\documentclass{JFM-FLM_Au}

\lefttitle{D. Nishiyama and Y. Hattori}
\righttitle{Journal of Fluid Mechanics}

\title{Nonlinear dynamics of short-wave instabilities in a helical vortex within and beyond the weakly nonlinear regime}

\author{Daisuke Nishiyama\aff{1,2} \and Yuji Hattori\aff{2}}

\affiliation{\aff{1}Graduate School of Information Sciences, Tohoku University, Sendai 980-8579, Japan
\aff{2}Institute of Fluid Science, Tohoku Universty, Sendai 980-8577, Japan}

\corresau{Daisuke Nishiyama, \email{nishiyama.daisuke.s2@dc.tohoku.ac.jp}}

\begin{document}
\maketitle

\begin{abstract}
The nonlinear evolution of a helical vortex with axial flow disturbed by an elliptic or curvature instability mode is investigated by direct numerical simulations. The normalized vortex core radius and reduced pitch are systematically varied to examine a wide range of linear growth rates, revealing two types of nonlinear evolution. At relatively small growth rates, both instabilities exhibit weakly nonlinear saturation, with the vortex and unstable mode structures largely preserved. Together with previous findings for other vortex configurations, these results reveal that weakly nonlinear saturation is a generic feature of short-wave instabilities. At larger growth rates, the instabilities reach higher energy levels and evolve beyond the weakly nonlinear regime, with the vortex and unstable mode structures substantially deformed. Comparison with weakly nonlinear analysis suggests that larger linear growth rates require larger disturbance amplitudes for nonlinear effects to balance linear growth. At these larger amplitudes, higher-order nonlinear effects and modal components not retained in the weakly nonlinear analysis become increasingly important, as demonstrated by the numerical results. This provides a new physical interpretation of the departure from the weakly nonlinear regime. Since helical vortices serve as a model for tip vortices shed from rotating blades, these results offer useful insights for tip-vortex control in wind turbines, which can potentially accelerate wake recovery and thereby improve turbine performance.
\end{abstract}



\section{Introduction}
Vortices are fundamental structures in fluid flows across a wide range of scales.
Their dynamics, particularly their stability and breakdown, have therefore attracted considerable attention for many decades.
For example, straight vortices can provide a useful framework for understanding phenomena ranging from aircraft trailing vortices \citep{breitsamter_wake_2011} to small-scale interacting vortices involved in turbulence generation \citep{mckeown_turbulence_2020}.
Natural extension of a straight vortex includes a vortex ring, characterized by curvature, and a helical vortex, characterized by both curvature and torsion.
These configurations have been studied as canonical models for understanding fundamental vortex characteristics \citep{saffman_vortex_1992, hardin_velocity_1982}.
More recently, helical vortices have attracted increasing attention because of their relevance to wind-turbine wakes, where coherent helical tip vortices can strongly influence wake recovery \citep{posa_instability_2021, meyers_wind_2022}.
Their destabilization and breakdown therefore provide a route to enhance wake mixing and reduce the velocity deficit affecting the performance of downstream turbines \citep{abraham_experimental_2023, brown_accelerated_2022}.
This motivates clarifying the mechanisms of turbulent transition in a helical vortex and exploiting this understanding for wake control.

Instabilities of a helical vortex can be classified into long- and short-wave instabilities according to the disturbance wavelength relative to the vortex-core size \citep{leweke_long-_2014}.
The long-wave instability is associated with the global structure of the helical vortex \citep{widnall_stability_1972,quaranta_long-wave_2015}.
In contrast, the short-wave instability is associated with the local structure of the vortex core and can therefore occur in vortices with different global geometries but similar core structures.
Representative examples are the elliptic instability, induced by strain acting on the vortex core \citep{kerswell_elliptical_2002}, and the curvature instability, induced by curvature of the vortex axis \citep{hattori_short-wavelength_2003, fukumoto_curvature_2005}.
Both arise through resonant interactions involving Kelvin waves, and these mechanisms can be generalized to multipolar strain \citep{le_dizes_short-wavelength_1999,ayapilla_triangular_2025,ayapilla_triangular_2026}.
Since a helical vortex is subject to both strain and curvature \citep{blanco-rodriguez_internal_2015}, the elliptic and curvature instabilities have been identified theoretically and numerically in helical vortices \citep{hattori_modal_2014,blanco-rodriguez_elliptic_2016,blanco-rodriguez_curvature_2017,xu_numerical_2025}.

One plausible route to turbulence is the growth of a dominant linear unstable mode to finite amplitude, followed by nonlinear evolution \citep{schmid_stability_2001}.
Therefore, understanding this nonlinear regime is important for clarifying the transition process.
For the long-wave instability of a helical vortex, characteristic nonlinear behaviors, such as leapfrogging and vortex-ring ejection \citep{quaranta_local_2019, hattori_effect_2024}, have been reported. 
In contrast, numerical and experimental studies have reported the short-wave instability during the nonlinear evolution of helical vortices \citep{leweke_long-_2014, kleine_tip-vortex_2019}, but its nonlinear dynamics has not been systematically investigated.

Nevertheless, because the short-wave instability is governed primarily by the local structure of the vortex core, studies of other vortex configurations provide insight into common features of its nonlinear dynamics.
Among these configurations, the elliptic instability $(-1,1)$ of strained columnar vortices has been extensively studied \citep{leweke_dynamics_2016}.
Here, the two integers denote the azimuthal wavenumbers of the resonant Kelvin waves.
From a theoretical perspective, weakly nonlinear analysis was developed for this configuration, predicting that nonlinear effects can saturate the growth of the elliptic instability $(-1,1)$ at finite amplitude, without leading directly to turbulent transition \citep{sipp_weakly_2000}. 
Such weakly nonlinear saturation was observed experimentally and numerically \citep{eloy_elliptic_2003, schaeffer_nonlinear_2010}.
In contrast, some numerical studies have also reported strongly nonlinear evolution involving substantial deformation or breakdown of the vortex \citep{laporte_direct_2000, tunkeaw_effect_2018}.
These results suggest two different types of nonlinear evolution: one remaining within a weakly nonlinear regime and the other evolving beyond it.
A previous study has indicated that the type of evolution depends on the parameter regime \citep{schaeffer_nonlinear_2010}.

Several important questions remain regarding the nonlinear evolution of short-wave instabilities.
First, it remains unclear whether weakly nonlinear dynamics leading to saturation is generic to short-wave instabilities.
Of particular interest in recent studies are instabilities of vortices with Gaussian-type profiles and axial flow \citep{brynjell-rahkola_numerical_2020, xu_numerical_2025}, such as the elliptic instability $(-2,0)$ and the curvature instability $(-1,0)$.
Axial flow modifies the dispersion relationships of Kelvin waves and can change the dominant resonant mode pairs, while also being commonly present in practical vortex flows \citep{leweke_dynamics_2016}.
For such instabilities, conventional weakly nonlinear analysis is not always straightforward because the presence and influence of critical layers depend on the Kelvin-wave branches involved.
Clear numerical and experimental evidence is also limited, particularly because viscous evolution of the base flow, including the changes in the axial flow, can also detune the resonance condition and obscure nonlinear saturation \citep{sipp_widnall_2003}.

Second, the physical mechanisms governing the departure from weakly nonlinear dynamics require further clarification.
In some confined configurations, such as an elliptic cylinder, the saturated state of the primary instability can itself be unstable through triadic interactions with additional modes \citep{mason_nonlinear_1999, kerswell_secondary_1999}, providing one possible route to more strongly nonlinear dynamics.
For an unconfined vortex with a smooth vorticity profile, however, it is less clear whether the same mechanism operates.
Indeed, for a strained Lamb-Oseen vortex, strongly nonlinear dynamics have been observed to develop directly from a single elliptic-instability mode \citep{schaeffer_nonlinear_2010}.
These observations suggest that further investigation of the departure from weakly nonlinear dynamics is still needed.
Clarifying this departure requires an extensive investigation of the parameter dependence of the nonlinear evolution, together with comparison with weakly nonlinear analysis. 

Motivated by the engineering relevance of helical vortices and the unresolved questions outlined above, the present study investigates, by direct numerical simulations, the nonlinear evolution of the elliptic and curvature instabilities in a helical vortex with a Gaussian core profile and axial flow.
The first objective is to determine whether weakly nonlinear dynamics leading to saturation is a common feature of these short-wave instabilities.
The second is to clarify what governs the departure from weakly nonlinear dynamics and how this departure occurs, both physically and in terms of its parameter dependence.
Addressing these questions provides useful insights for tip-vortex control in wind turbines, while also advancing the understanding of the nonlinear dynamics of short-wave instabilities.

This paper is organized as follows.
In \S~\ref{sec:outline}, we describe the numerical procedure.
In \S~\ref{sec:results}, the results of the direct numerical simulations are presented.
These results are further discussed in \S~\ref{sec:discussion} through comparison with weakly nonlinear analysis.
We conclude in \S~\ref{sec:conclusion}.

\section{Numerical procedure}\label{sec:outline}
\subsection{Outline}
\begin{figure}
  \centerline{\includegraphics[scale=1.2]{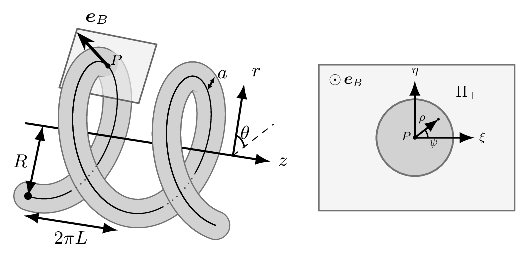}}
  \caption{Schematic of the problem configuration and coordinate systems.}
\label{fig:schematic}
\end{figure}

We investigate the nonlinear evolution of a helical vortex disturbed by a short-wave instability in an incompressible viscous fluid.
A schematic of the problem configuration is shown in figure~\ref{fig:schematic}.
The base flow is a helical vortex with constant pitch $2{\pi}L$.
It is described in cylindrical coordinates $(r,\theta,z)$ and is periodic in the $z$ direction with a period $2\pi L$.
Other parameters that determine the shape of the helical vortex are the helix radius $R$ and the core radius $a$.
These three length scales define two non-dimensional parameters:
the reduced pitch $L/R$ and the normalized core radius $\varepsilon=aR/(R^2+L^2)$.
The base flow and its short-wave instabilities are further characterized by the Reynolds number $Re=\Gamma/\nu$ and the axial flow parameter $W$, where $\Gamma$ and $\nu$ denote the circulation of the helical vortex and the kinematic viscosity, respectively.
The parameter $W$ is defined in \S~\ref{sec:base flow}.
Accordingly, the present problem is mainly characterized by the four non-dimensional parameters $L/R, \varepsilon, Re$, and $W$.

The numerical procedure of this study is as follows: 
(i) a helical vortex in a quasi-steady state is obtained as the base flow; 
(ii) an unstable mode corresponding to a short-wave instability of the base flow is obtained; 
and (iii) the base flow is disturbed by the short-wave instability mode, and the subsequent nonlinear evolution is investigated.
The Navier-Stokes equations are solved numerically in steps (i) and (iii), whereas the linearized Navier-Stokes equations about the base flow are solved numerically in step (ii).
The numerical methods used in these calculations follow those of \citet{hattori_numerical_2019} and are summarized in \S~\ref{sec:numerical methods}.
The procedures for steps (i) and (ii) have already been demonstrated \citep{xu_numerical_2025, hattori_effect_2024} and are briefly summarized in \S~\ref{sec:base flow} and \S~\ref{sec:linear instability}, respectively.
Step (iii) constitutes the main focus of the present study; its numerical procedure and simulation parameters are described in \S~\ref{sec:nonlinear evolution} and \S~\ref{sec:simulation parameters}, respectively.

\subsection{Numerical methods}\label{sec:numerical methods}
The incompressible Navier-Stokes equations in cylindrical coordinates are solved numerically using the method described in \citet{hattori_numerical_2019}.
The same spatial and temporal discretizations are applied to the linearized Navier-Stokes equations.
For the spatial discretization, the sixth-order accurate compact scheme is used in the $r$ direction \citep{lele_compact_1992}, while the Fourier spectral method is used in the $\theta$ and $z$ directions.
The numerical domain is $0\leq r \leq R_{far}=100R$, $0\leq \theta < 2\pi$, and $0\leq z < 2\pi L$, where $R_{far}$ denotes the radial location of the outer boundary.
The outer radial boundary is placed sufficiently far from the helical vortex that its effect on the flow is negligible, and slip boundary conditions are imposed there.
In the $\theta$ and $z$ directions, periodic boundary conditions are imposed.
For the temporal discretization, the Crank-Nicolson scheme is used for the viscous terms, while the second-order Adams-Bashforth method is used for the other terms. 
For details of the discretized equations, see, for example, appendix E of \citet{xu_numerical_2025}.
A non-uniform stretched grid is used in the $r$ direction \citep{hattori_numerical_2019}.
The radial grid spacing is $\Delta r\approx 7.14\times10^{-3} R$ for $r\lesssim 1.5R$, including the region surrounding the helical vortex, and is gradually increased towards the outer boundary, reaching $\Delta r\approx 2R$ at $r=R_{far}$.
The number of grid points or Fourier modes is $G_r\times G_\theta\times G_z=565\times 896\times 256$ for $L/R=0.3$, while $G_z$ is replaced with $448$ for $L/R=0.5$ and $0.55$.
The values of $G_\theta$ and $G_z$ are chosen such that the corresponding physical grid spacings near the helical vortex are approximately equal to the minimum radial grid spacing.
The time step is set to $\Delta t=7.5\times10^{-4} R^2/\Gamma$.
The adequacy of the spatial resolution and time step is examined in appendix~\ref{appA}.
In the following, all variables are nondimensionalized using $R$ and $\Gamma$, with $R$ and $R^2/\Gamma$ taken as the length and time scales, respectively, unless stated explicitly \citep{hattori_effect_2024, xu_numerical_2025}.

\subsection{Base flow}\label{sec:base flow}

\begin{figure}
  \centerline{\includegraphics[scale=.35]{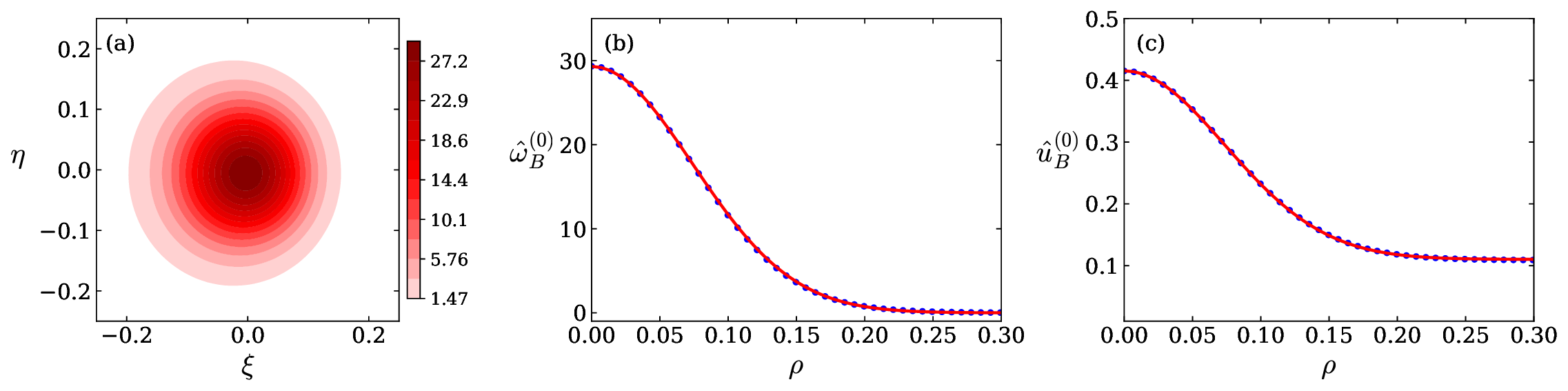}}
  \caption{Local structure of the base flow in a quasi-steady state for $L/R=0.3, \varepsilon=0.1, Re=1.57\times10^4$, and $W=0.2$: (a)contours of vorticity component $\omega_B$; radial distribution of (b)$\hat{\omega}^{(0)}_B$ and (c)$\hat{u}^{(0)}_B$. Circles and solid lines in (b,c) show the numerical results and the fitted Gaussian profiles, respectively.}
\label{fig:baseflow}
\end{figure}

We consider a helical vortex for which the vorticity and velocity components tangent to the vortex centerline have Gaussian profiles.
This configuration has been widely investigated in previous studies \citep{brynjell-rahkola_numerical_2020, xu_numerical_2025}.
Although an analytical solution for a quasi-steady state is not available, it can be obtained numerically \citep{selcuk_helical_2017}.
The initial condition is constructed by prescribing profiles similar to those of a Batchelor vortex around the centerline helix.
Because this initial field is not an exact equilibrium solution, the Navier-Stokes equations are integrated to allow the vortex to relax to a quasi-steady state.
After the relaxation process, the helical vortex propagates uniformly owing to its self-induced velocity and is quasi-steady in a suitably translating or rotating frame.
Here, we adopt a frame translating along the $z$ axis as in \citet{hattori_effect_2024}, unless otherwise stated.

Because short-wave instabilities are sensitive to the detailed structure of the base flow in the vicinity of the vortex core, we introduce a local coordinate system in a plane perpendicular to the centerline helix, as commonly done in previous theoretical and numerical studies \citep{blanco-rodriguez_internal_2015, xu_numerical_2025}.
Let $P=(r_P,\theta_P,z_P)$ be a point on the centerline helix.
A local orthonormal basis $(\boldsymbol{e}_{\xi}, \boldsymbol{e}_{\eta}, \boldsymbol{e}_B)$, related to the Frenet-Serret frame at $P$, is defined by 

\begin{equation}
\boldsymbol{e}_{\xi} = \boldsymbol{e}_r(\theta_P),
  \quad
\boldsymbol{e}_{B} = \left(1+\frac{r^2_P}{L^2} \right)^{-1/2} \left \lbrack \frac{r_P}{L}\boldsymbol{e}_{\theta}(\theta_P)+\boldsymbol{e}_z \right \rbrack,
  \quad
\boldsymbol{e}_{\eta} = \boldsymbol{e}_B\times \boldsymbol{e}_{\xi},
\end{equation}
where $\boldsymbol{e}_r,\boldsymbol{e}_\theta$ and $\boldsymbol{e}_z$ are the unit basis vectors of the cylindrical coordinate system.
We denote by $\Pi_{\perp}$ the plane passing through $P$ and perpendicular to $\boldsymbol{e}_B$ (figure \ref{fig:schematic}); $\boldsymbol{e}_\xi$ and $\boldsymbol{e}_\eta$ form an orthonormal basis for this plane.
The local Cartesian coordinates $(\xi,\eta)$ are defined on $\Pi_\perp$, with the origin at the point $P$ and the basis $(\boldsymbol{e}_\xi, \boldsymbol{e}_\eta)$.
We also introduce the corresponding local polar coordinates $(\rho, \psi)$ on $\Pi_\perp$.
The associated basis $(\boldsymbol{e}_\rho, \boldsymbol{e}_\psi)$ is defined as

\begin{equation}
\boldsymbol{e}_{\rho} = \cos{\psi} \boldsymbol{e}_\xi + \sin{\psi}\boldsymbol{e}_\eta,
  \quad
\boldsymbol{e}_{\psi} = -\sin{\psi} \boldsymbol{e}_\xi + \cos{\psi}\boldsymbol{e}_\eta.
\end{equation}
Here, $\rho$ denotes the distance from the point $P$, and $\psi$ is the angle measured from the $\boldsymbol{e}_\xi$ direction.
It should be noted that the local structure of the base flow does not depend on the choice of $P$ due to the helical symmetry \citep{selcuk_helical_2017}.

Using these local coordinates, the resulting quasi-steady helical vortex is shown in figure \ref{fig:baseflow}.
Figure \ref{fig:baseflow}(a) shows the vorticity component in the $\boldsymbol{e}_B$ direction, denoted by $\omega_B$.
The corresponding velocity component is denoted by $u_B$.
It can be seen that the vorticity distribution in the vortex core is slightly deformed from an axisymmetric Gaussian profile owing to the self-induced strain.
To quantify this non-axisymmetric deformation and to define the core radius $a$ and axial flow parameter $W$, the vorticity and velocity components $\omega_B$ and $u_B$ are decomposed as  

\begin{equation}
\omega_B(\rho,\psi) =\sum\limits_{m} \hat{\omega}^{(m)}_B(\rho) \exp{(\mathrm{i}m\psi)},
  \quad
u_B(\rho,\psi) =\sum\limits_{m} \hat{u}^{(m)}_B(\rho) \exp{(\mathrm{i}m\psi)},
\end{equation}
where $ \hat{\omega}^{(m)}_B$ and $\hat{u}^{(m)}_B$ denote the Fourier coefficients.
In the quasi-steady state, the axisymmetric components $\hat{\omega}^{(0)}_B$ and $\hat{u}^{(0)}_B$ approximately remain close to the Gaussian profiles \citep{xu_numerical_2025}

\begin{equation}\label{eq:gaussian}
\hat{\omega}^{(0)}_{B}(\rho)\approx\frac{\Gamma}{\pi{a^2}}\exp{\left( -\frac{\rho^2}{a^2} \right)},
  \quad
\hat{u}^{(0)}_{B}(\rho){\approx}W_{0}\exp{\left( -\frac{\rho^2}{a^2} \right)}+W_{00},
\end{equation}
where $W_0$ denotes the amplitude of the axial flow and $W_{00}$ denotes the uniform velocity associated with the translating reference frame.
The axial flow parameter $W$ is defined by normalizing $W_0$ by the characteristic velocity in the vortex core, $\Gamma/(2\pi a)$ \citep{xu_quasi-equilibrium_2022}. 
Therefore, the vortex core $a$ and axial flow parameter $W$ are determined by the best fit of $\hat{\omega}^{(0)}_{B}$ and $\hat{u}^{(0)}_{B}$ to equation~\eqref{eq:gaussian}.
Figures~\ref{fig:baseflow}(b) and (c) show $\hat{\omega}^{(0)}_B$ and $\hat{u}^{(0)}_B$, respectively, together with the corresponding fitted profiles. 
Their close agreement with the Gaussian profiles in equation~\eqref{eq:gaussian} shows that the local core structure of the quasi-steady helical vortex is close to that of a Batchelor vortex.
We note that the short-wave instabilities considered below are driven by non-axisymmetric components $(m\neq0)$ of the base flow.

\subsection{Linear instability}\label{sec:linear instability}

\begin{figure}
  \centerline{\includegraphics[scale=.42]{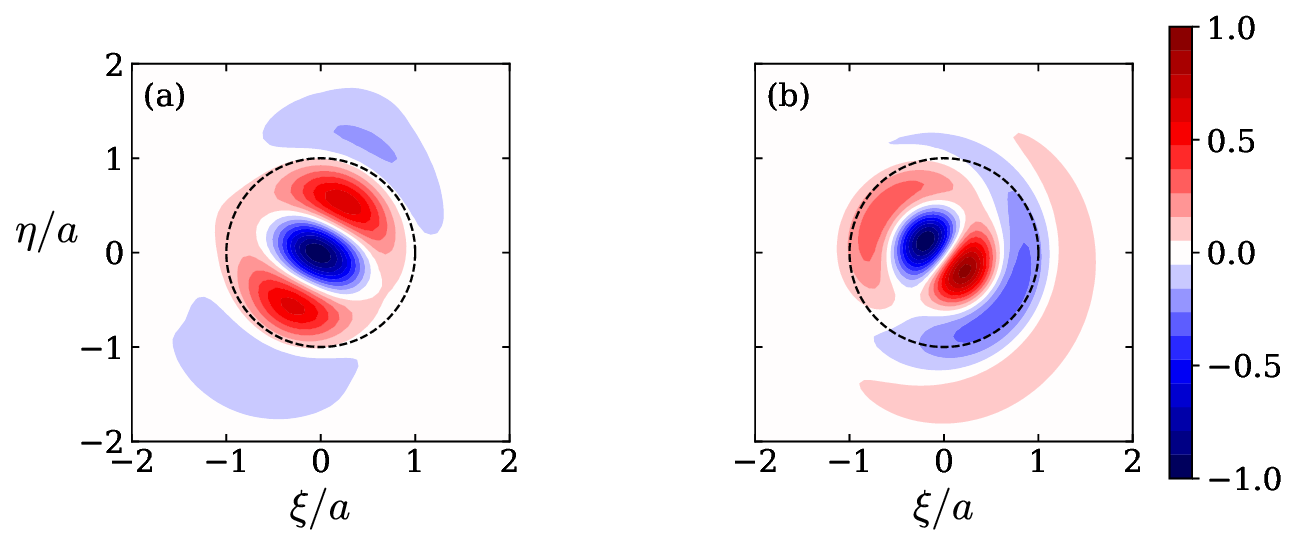}}
  \caption{Contours of $\omega'_{B,N}$ normalized by its maximum absolute value for (a) the elliptic instability $(-2,0;[2,2])$ and (b) the curvature instability $(-1,0;[2,4])$ in the local coordinates. The dashed black circles indicate $\rho=a$.}
\label{fig:linear}
\end{figure}

The short-wave instabilities of the helical vortex are investigated using the base flow obtained in \S ~\ref{sec:base flow}.
The basic idea of the linear stability analysis is the same as that used for a vortex ring \citep{hattori_numerical_2019}: the Navier-Stokes equations linearized about the base flow are integrated numerically in a cylindrical coordinate system.
For a helical vortex, however, the helical symmetry leads to a different modal decomposition of the disturbance.

Owing to the helical symmetry, the cylindrical components of the base-flow velocity depend only on $r$ and $\varphi=\theta-z/L$ \citep{selcuk_helical_2017}.
The total velocity field $\boldsymbol{u}$ is then decomposed into the base flow and an initially small disturbance as

\begin{equation}
\boldsymbol{u}(r,\theta,z,t) = \boldsymbol{u}^{\mathrm{BF}}(r,\varphi)+ \boldsymbol{u}'(r,\varphi,z,t),
\end{equation}
where $\boldsymbol{u}^{\mathrm{BF}}$ and $\boldsymbol{u}'$ denote the base-flow and disturbance velocity fields, respectively.
As commonly assumed in previous stability analyses \citep{brynjell-rahkola_numerical_2020,hattori_numerical_2019}, the base flow is frozen in time during the linear stability analysis, and its viscous evolution is neglected.
Owing to this reduced dependence, disturbances of the following form evolve independently under the linearized equations:

\begin{equation}\label{eq:mode}
\boldsymbol{u}'_{N}(r,\varphi,z,t) = \tilde{\boldsymbol{u}}'_N(r,\varphi,t)\exp{(\mathrm{i} \frac{Nz}{L})}+\mathrm{c.c.},
\end{equation}
where $N$ is a positive integer specifying the mode, $\tilde{\boldsymbol{u}}'_N$ is the corresponding mode function, and $\mathrm{c.c.}$ denotes the complex-conjugate term.
Expanding $\tilde{\boldsymbol{u}}'_N$ in a Fourier series with respect to $\varphi$ \citep{xu_numerical_2025} and using $\varphi=\theta-z/L$, equation \eqref{eq:mode} can then be written in cylindrical coordinates as

\begin{equation}
\boldsymbol{u}'_{N}(r,\theta,z,t) = \sum\limits_{l} \hat{\boldsymbol{u}}'_{l,N-l} (r,t)\exp{(il\theta)} \exp{\left[ i\frac{(N-l)z}{L} \right]}+\mathrm{c.c.}.
\end{equation}
Therefore, when the disturbance velocity is decomposed in cylindrical coordinates as

\begin{equation}
\boldsymbol{u}'(r,\theta,z,t) = \sum\limits_{l}\sum\limits_{k} \hat{\boldsymbol{u}}'_{l,k} (r,t)\exp{(il\theta)} \exp{\left( i\frac{kz}{L} \right)},
\end{equation}
the disturbance $\boldsymbol{u}'_{N}$ is obtained from the components that satisfy $l+k=\pm N$.
This allows the independently evolving mode defined in equation~\eqref{eq:mode} to be obtained within the cylindrical-coordinate formulation.

To obtain the most unstable eigenmode for a given $N$, the initial disturbance of $\boldsymbol{u}'_N$ is prescribed as a random velocity field localized around the vortex core.
The linearized Navier--Stokes equations are then integrated for a sufficiently long time, allowing the disturbance to converge to the most unstable mode.
In the present study, we focus on the elliptic instability mode $(-2,0;[2,2])$ and the curvature instability mode $(-1,0;[2,4])$, both of which are among the dominant short-wave instabilities under the axial-flow condition considered here \citep{blanco-rodriguez_elliptic_2016, blanco-rodriguez_curvature_2017}.
In this notation, the final labels $[2,2]$ and $[2,4]$ identify the Kelvin modes involved in the respective instabilities; see, for example, \citet{leweke_dynamics_2016, hattori_numerical_2019} for details.
The resulting mode structures are shown in figure \ref{fig:linear}.
Hereafter, $u'_{B,N}$ and $\omega'_{B,N}$ denote the $\boldsymbol{e}_B$-component of the velocity and vorticity disturbances of mode $N$, respectively.
When represented in the local coordinates, the mode structures are in good qualitative agreement with those reported in previous studies \citep{xu_numerical_2025, hattori_numerical_2019}.

\subsection{Nonlinear evolution}\label{sec:nonlinear evolution}
The nonlinear evolution of each short-wave instability is investigated using the base flow and unstable mode obtained in \S~\ref{sec:base flow} and \S~\ref{sec:linear instability}, respectively.
The initial velocity field consists of the base flow of the helical vortex and a small-amplitude disturbance corresponding to a single short-wave instability mode, which can be written as $\boldsymbol{u}=\boldsymbol{u}^{\mathrm{BF}}+ \boldsymbol{u}'_N$.
The initial disturbance amplitude is chosen to be sufficiently small such that the disturbance initially grows exponentially as predicted in \S~\ref{sec:linear instability}.
As in the linear stability analysis, the base flow is frozen in time during the nonlinear evolution, following the approach adopted for the nonlinear analysis of the elliptic instability in a strained columnar vortex \citep{schaeffer_nonlinear_2010}.
This provides a natural first step towards extending the linear analysis into the nonlinear regime, while isolating the nonlinear disturbance dynamics from the viscous evolution of the base flow.
In addition, this treatment enables a direct comparison with the weakly nonlinear analysis.
Its validity is examined in appendix~\ref{appB}.
During the nonlinear evolution, higher harmonic components, such as $\boldsymbol{u}'_{2N}$ and $\boldsymbol{u}'_{3N}$, are generated through nonlinear interactions involving $\boldsymbol{u}'_{N}$ and the subsequently generated harmonics.
Although these harmonic components are allowed to develop, non-harmonic disturbance components are not generated, since only a single instability mode is introduced as the initial disturbance.
Consequently, triadic interactions involving non-harmonic disturbance components, such as those reported by \citet{mason_nonlinear_1999}, are not considered here.
Such interactions may lead to more complex dynamics and are beyond the scope of the present study, which focuses on the nonlinear evolution initiated by a single short-wave instability mode.

\subsection{Simulation parameters}\label{sec:simulation parameters}

\begin{table}
  \begin{center}
  \begin{tabular}{cc}
    \begin{minipage}[t]{0.47\linewidth}
      \begin{center}
      Elliptic instability $(-2,0;[2,2])$ \\[4pt]
      \begin{tabular}{cccccc}
        $Re$ & $W$ & $\varepsilon$ & $L/R$ & $N$ & $\sigma$ \\[3pt]
        $1.57\times10^{4}$ & $0.20$ & $0.10$ & $0.30$ & $32$ & $6.59\times 10^{-2}$ \\
        $1.57\times10^{4}$ & $0.20$ & $0.15$ & $0.30$ & $21$ & $1.45\times 10^{-1}$ \\
        $1.57\times10^{4}$ & $0.20$ & $0.20$ & $0.30$ & $15$ & $1.67\times 10^{-1}$ \\
        $1.57\times10^{4}$ & $0.20$ & $0.15$ & $0.20$ & $21$ & $3.78\times 10^{-1}$ \\
        $1.57\times10^{4}$ & $0.20$ & $0.15$ & $0.55$ & $18$ & $2.26\times 10^{-2}$ \\
      \end{tabular}
      \end{center}
    \end{minipage}
    &
    \begin{minipage}[t]{0.47\linewidth}
      \begin{center}
      Curvature instability $(-1,0;[2,4])$\\[4pt]
      \begin{tabular}{cccccc}
        $Re$ & $W$ & $\varepsilon$ & $L/R$ & $N$ & $\sigma$ \\[3pt]
        $2.24\times10^{4}$ & $0.20$ & $0.11$ & $0.50$ & $15$ & $1.31\times 10^{-2}$ \\
        $2.24\times10^{4}$ & $0.20$ & $0.13$ & $0.50$ & $13$ & $1.83\times 10^{-2}$ \\
        $2.24\times10^{4}$ & $0.20$ & $0.16$ & $0.50$ & $10$ & $2.95\times 10^{-2}$ \\
      \end{tabular}
      \end{center}
    \end{minipage}
  \end{tabular}
  \caption{Parameters used for the nonlinear simulations of the elliptic and the curvature instability cases.}
  \label{tab:parameter}
  \end{center}
\end{table}

As described in \S~\ref{sec:numerical methods}, the present problem is mainly characterized by the non-dimensional parameters $L/R$, $\varepsilon$, $Re$, and $W$.
For each parameter set, the mode number $N$ is chosen such that the corresponding disturbance wavenumber is close to the resonance condition associated with the maximum growth rate; see \citet{xu_numerical_2025, brynjell-rahkola_numerical_2020} for the choice of $N$.
The parameter sets considered in the nonlinear simulations are summarized in table~\ref{tab:parameter}.
A relatively high Reynolds number of order $10^4$ is used \citep{leweke_long-_2014}, and the axial flow parameter is fixed at $W=0.2$ \citep{hattori_numerical_2019}, whereas the geometric parameters $\varepsilon$ and $L/R$ are varied to examine the nonlinear evolution over a range of linear growth rates $\sigma$.
For the elliptic instability, the growth rate increases with increasing $\varepsilon$ and decreasing $L/R$ within a typical parameter range \citep{blanco-rodriguez_elliptic_2016}.
The latter dependence arises because a smaller pitch brings neighboring turns of the helical vortex closer together, thereby increasing the strain imposed on the vortex core.
For the curvature instability, the growth rate increases with $\varepsilon$ and is independent of $L/R$ at leading order \citep{blanco-rodriguez_curvature_2017}.
Accordingly, $L/R$ is fixed and $\varepsilon$ is varied for the curvature-instability cases.

\section{Results}\label{sec:results}
\subsection{Overview of the nonlinear evolution}\label{sec:overview}

\begin{figure}
  \centerline{\includegraphics[scale=.38]{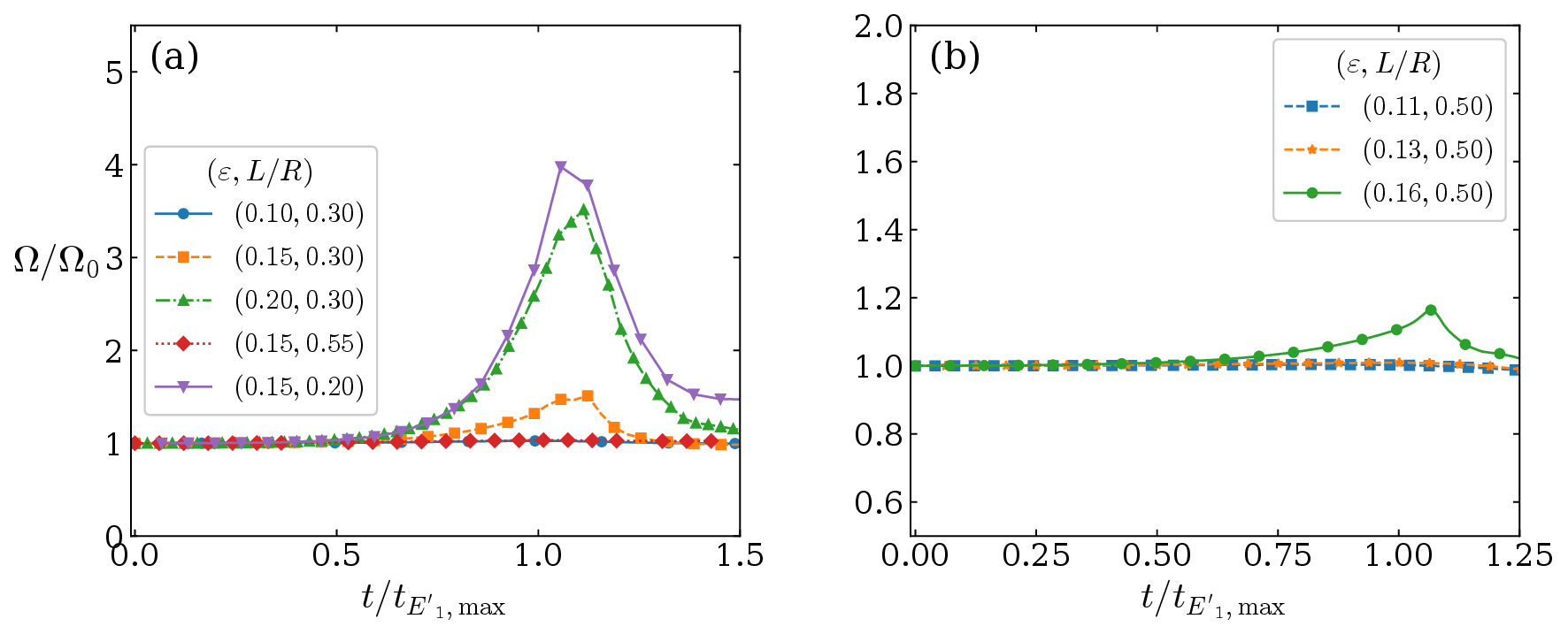}}
  \caption{Time evolution of the total enstrophy $\Omega$ normalized by its initial value $\Omega_0$ for (a) the elliptic instability $(-2,0;[2,2])$ and (b) the curvature instability $(-1,0;[2,4])$. Time is normalized by $t_{E'_1,\max}$, the time at which the energy of the unstable component $E'_1$ reaches its maximum over the simulated time interval.}
\label{fig:enstrophy}
\end{figure}

\begin{figure}
  \centerline{\includegraphics[scale=.32]{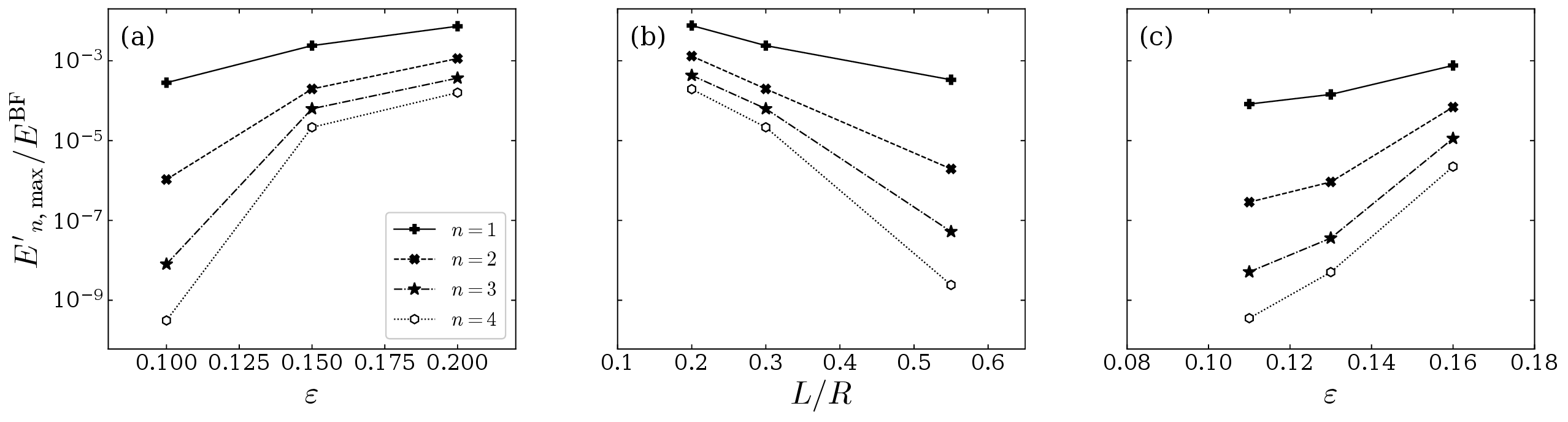}}
  \caption{Maximum normalized disturbance energies, $E'_{n,\max}/E^{\mathrm{BF}}$, attained over the simulated time interval for (a) the elliptic instability $(-2,0;[2,2])$ with $L/R=0.3$ and varying $\varepsilon$, (b)the elliptic instability with $\varepsilon=0.15$ and varying $L/R$, and (c) the curvature instability $(-1,0;[2,4])$ with $L/R=0.5$ and varying $\varepsilon$.}
\label{fig:energy_max}
\end{figure}

We first summarize the main features of the nonlinear evolution of the short-wave instabilities observed in the present study.
As shown in what follows, the resulting nonlinear evolution can be broadly classified into two types: evolution that remains within the weakly nonlinear regime and evolution that proceeds beyond it.
To characterize the nonlinear evolution quantitatively, we introduce the disturbance energy and enstrophy.
The disturbance energy $E'_n$ associated with $\boldsymbol{u}'_{nN}$ and the base-flow energy $E^{\mathrm{BF}}$ are defined, respectively, as

\begin{equation}
E'_n=\int_V |\boldsymbol{u}'_{nN}|^2\,dV,
  \quad
E^{\mathrm{BF}}=\int_V |\boldsymbol{u}^{\mathrm{BF}}|^2\,dV.
\end{equation}
Here, $n=0, 1,2,\dots$, with $n=1$ corresponding to the initially imposed unstable mode and $n \ge 2$ to its higher harmonics; $n=0$ represents the mean-flow correction \citep{schaeffer_nonlinear_2010}.
The total enstrophy is defined as 

\begin{equation}
    \Omega= \int_V |\boldsymbol{\omega}|^2\,dV, 
\end{equation}
where $dV=r\,dr\,d\theta\,dz$ denotes the volume element in cylindrical coordinates and $\boldsymbol{\omega}$ is the total vorticity field. 

Figure \ref{fig:enstrophy} shows the time evolution of the total enstrophy normalized by its initial value $\Omega_0$.
Overall, the total enstrophy increases in cases with relatively large linear growth rates, such as $(\varepsilon, L/R)=(0.2,0.3)$ for the elliptic instability and $(\varepsilon,L/R)=(0.16,0.5)$ for the curvature instability.
This increase indicates a significant modification of the initial vortex structure, possibly involving the generation of smaller-scale structures.
In these cases, the enstrophy reaches its peak near the time at which $E'_1$ attains its maximum, suggesting that these structural changes are associated with the growth and nonlinear development of the initially imposed unstable component.
In contrast, for cases with relatively small growth rates, such as $(\varepsilon, L/R)=(0.1,0.3)$ for the elliptic instability and $(\varepsilon,L/R)=(0.11,0.5)$ for the curvature instability, no significant increase in the total enstrophy is observed, even near the maximum of $E'_1$.
This suggests that the vortex largely retains its initial structure in these cases.

Figure \ref{fig:energy_max} shows the maximum energies of the disturbance components $E'_{n,\max}$ for $n=1,2,3,$ and $4$, where $E'_{n,\max}$ denotes the maximum value of $E'_n$ attained during the simulation.
The values shown are normalized by the corresponding base-flow energy $E^{\mathrm{BF}}$ to allow a consistent comparison among cases with different base flows and to express the disturbance magnitude relative to the underlying base flow.
Overall, cases with larger linear growth rates tend to attain larger normalized values of $E'_{1,\max}$, indicating that the unstable component grows to a larger magnitude relative to the base flow.
This trend is observed for both parameter variations of the elliptic instability and for the curvature instability, as shown in figure~\ref{fig:energy_max}(a--c).
This suggests that nonlinear effects become more pronounced as the linear growth rate increases, since the nonlinear disturbance terms become increasingly important as the unstable component grows.
For example, for the elliptic instability at $(\varepsilon,L/R)=(0.2,0.3)$, the relatively large normalized value of $E'_{1,\max}$ is accompanied by appreciable energies in the higher harmonics, indicating relatively strong nonlinear interactions.
This behavior is also consistent with the observed increase in enstrophy.
In contrast, for the elliptic instability at $(\varepsilon,L/R)=(0.1,0.3)$, the disturbance components exhibit a clear energy hierarchy, with progressively smaller maximum energies at higher harmonics, indicating that nonlinear interactions remain relatively weak.

\begin{figure}
  \centerline{\includegraphics[scale=.35]{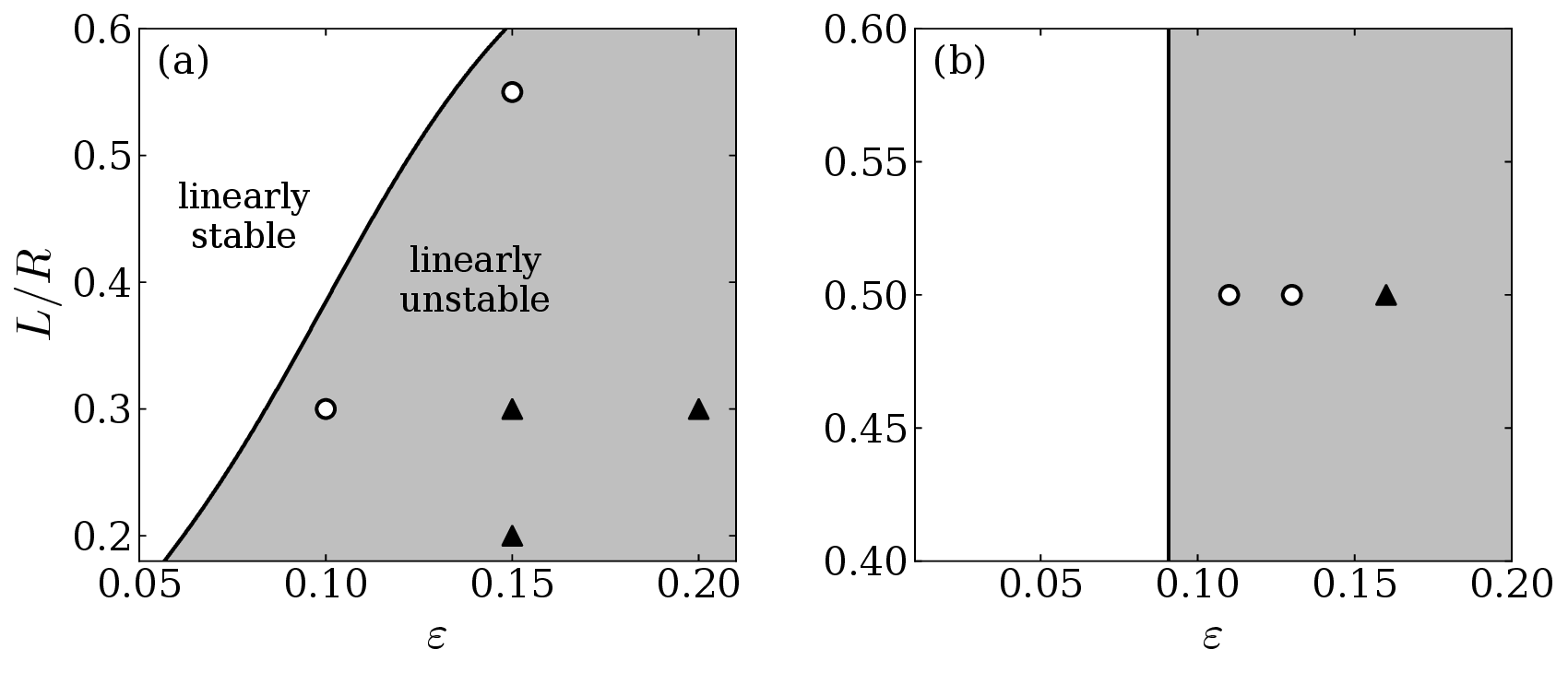}}
  \caption{Parameter combinations $(\varepsilon,L/R)$ and the corresponding types of nonlinear evolution for (a) the elliptic instability $(-2,0;[2,2])$ and (b) the curvature instability $(-1,0;[2,4])$. 
  Open circles and filled triangles denote cases whose nonlinear evolution remains within and evolves beyond the weakly nonlinear regime, respectively. 
  The solid black curves denote the neutral-stability boundaries based on the linear stability results of \cite{blanco-rodriguez_elliptic_2016,blanco-rodriguez_curvature_2017}.}
\label{fig:parameter}
\end{figure}

Based on the above discussion, the nonlinear evolution observed here can be broadly classified into two types: (i) evolution within the weakly nonlinear regime, in which the original vortex structure is largely preserved and the disturbance energy remains relatively low with only limited growth of higher harmonic components, and (ii) evolution beyond the weakly nonlinear regime, in which the original vortex structure is substantially deformed, possibly accompanied by the generation of small-scale structures, and the disturbance reaches higher energy levels with appreciable higher harmonic components.
The parameter combinations and the corresponding types of nonlinear evolution are summarized in figure~\ref{fig:parameter}.
These two types tend to be associated with relatively small and large linear growth rates, respectively, consistent with \citet{schaeffer_nonlinear_2010}.
Although the nonlinear evolution is broadly classified into these two types, its characteristics  appear to vary continuously between them as the linear growth rate increases.
This trend is discussed further in \S~\ref{sec:discussion} in comparison with the weakly nonlinear analysis.
In the following sections, these two types of nonlinear evolution are examined separately and in detail.

\subsection{Weakly nonlinear regime}
For both the elliptic and curvature instabilities, some cases remain within the weakly nonlinear regime throughout their nonlinear evolution. 
These cases are generally associated with relatively small linear growth rates.
In this section, we examine their dynamics in detail and show that the two instabilities exhibit common features characteristic of weakly nonlinear evolution, consistent with \citet{sipp_weakly_2000, schaeffer_nonlinear_2010}.

\subsubsection{Elliptic instability}\label{sec:WNL_elliptic}

\begin{figure}
  \centerline{\includegraphics[scale=.45]{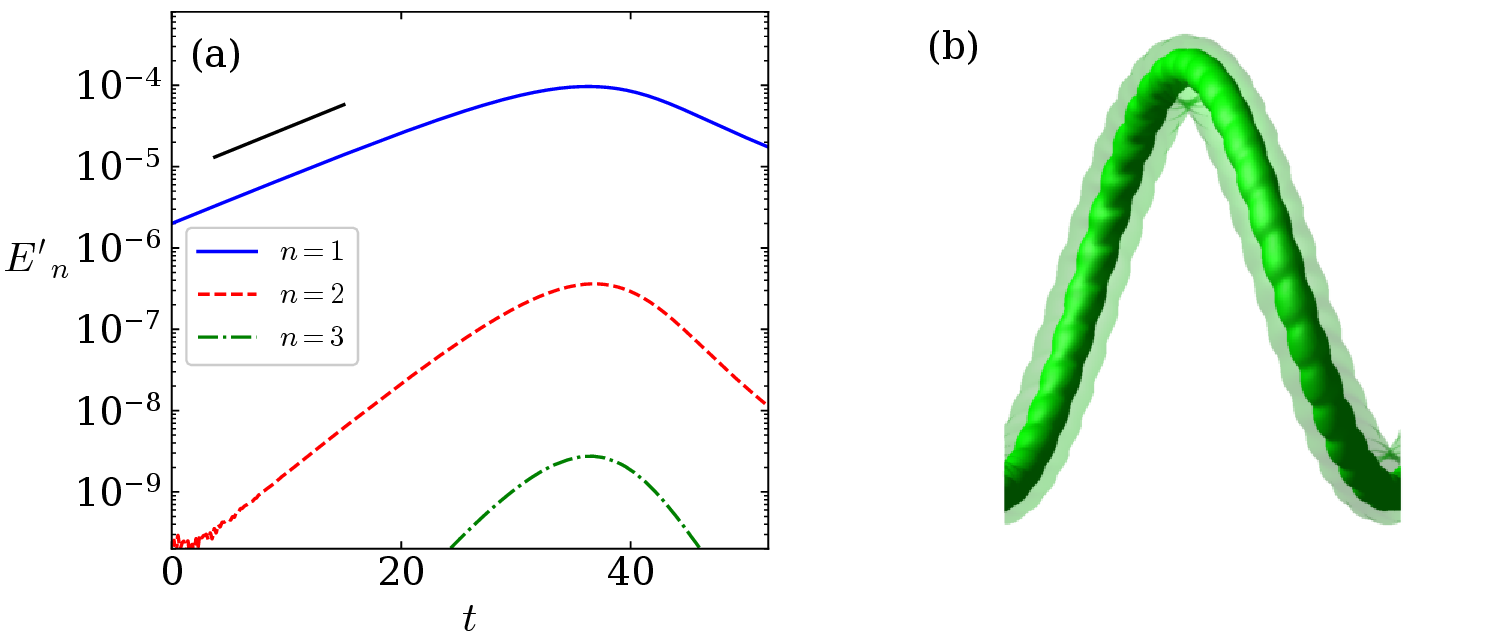}}
  \caption{Elliptic instability $(-2,0;[2,2])$ for $(\varepsilon,L/R)=(0.1,0.3)$. (a)Time evolution of the disturbance energies $E'_n$. The black line indicates the slope corresponding to the energy growth predicted by the linear stability analysis.
  (b)Isosurfaces of the vorticity magnitude $|\boldsymbol{\omega}|$ at $t=37.5$, when the energy of the unstable component is close to its maximum. The translucent and opaque isosurfaces correspond  to$|\boldsymbol{\omega}|/|\boldsymbol{\omega}_0|_{\max}=0.1$ and $|\boldsymbol{\omega}|/|\boldsymbol{\omega}_0|_{\max}=0.5$, respectively. $|\boldsymbol{\omega}_0|_{\max}$ denotes the maximum initial vorticity magnitude.} 
\label{fig:WNL_elliptic_energy}
\end{figure}

\begin{figure}
  \centerline{\includegraphics[scale=.25]{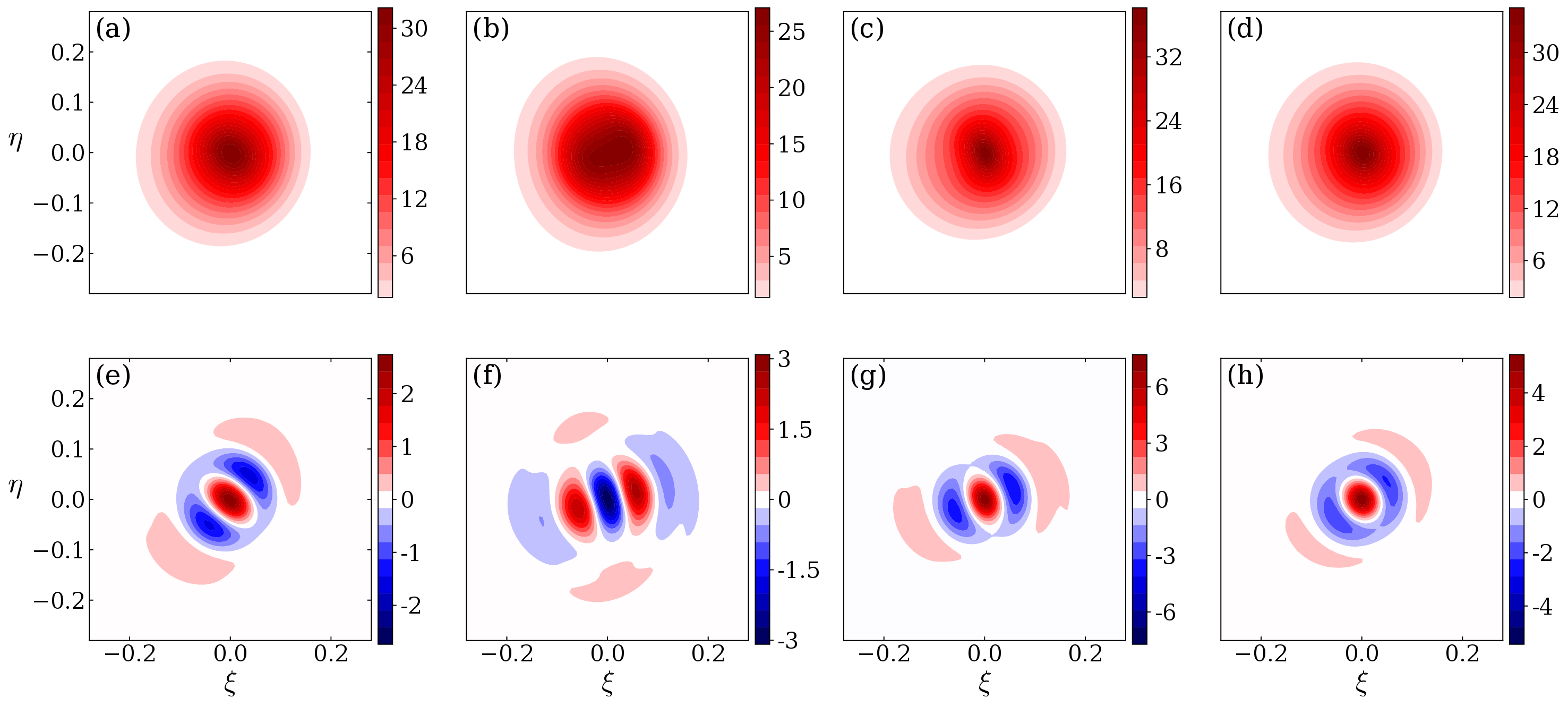}}
  \caption{Elliptic instability $(-2,0;[2,2])$ for $(\varepsilon,L/R)=(0.1,0.3)$. Contours of (a--d) the total vorticity $\omega_B$ and (e--h) the disturbance vorticity $\omega'_{B,N}$ in the local coordinates.
  From left to right: $t=13.5,25.5,37.5,49.5$.}
\label{fig:WNL_elliptic_bfield}
\end{figure}

\begin{figure}
  \centerline{\includegraphics[scale=.39]{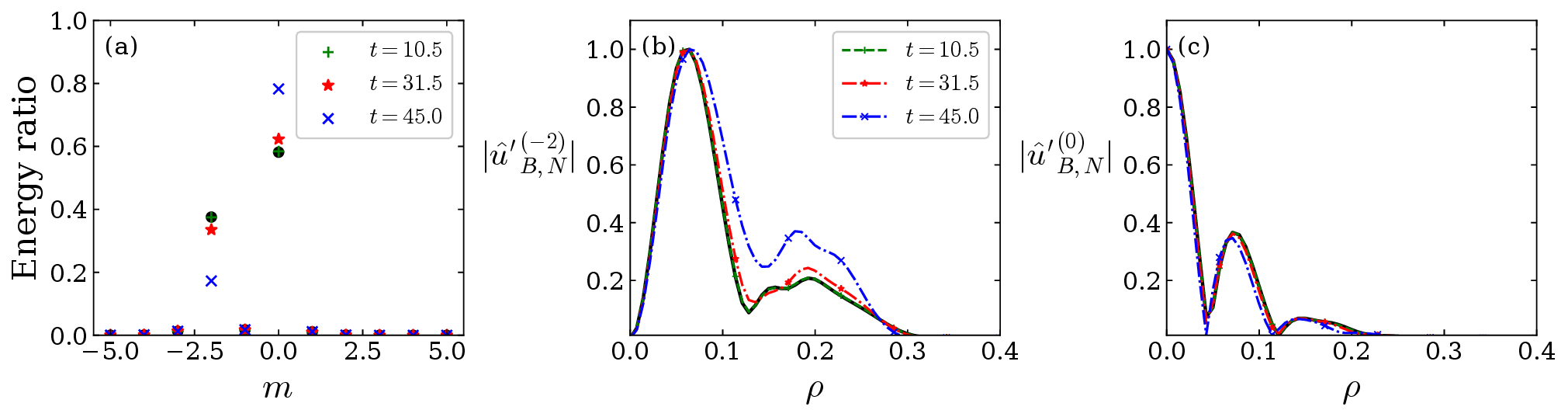}}
  \caption{Elliptic instability $(-2,0;[2,2])$ for $(\varepsilon,L/R)=(0.1,0.3)$. Structure of the unstable mode in the local coordinates: (a)energy ratios among the azimuthal components $\hat{u}_{B,N}^{\prime(m)}$ and radial distribution of $|\hat{u}_{B,N}^{\prime(m)}|$ for (b) $m=-2$ and (c) $m=0$. Black circles and solid lines denote the results of the linear stability analysis.}
\label{fig:WNL_elliptic_kelvin}
\end{figure}

\begin{figure}
  \centerline{\includegraphics[scale=.41]{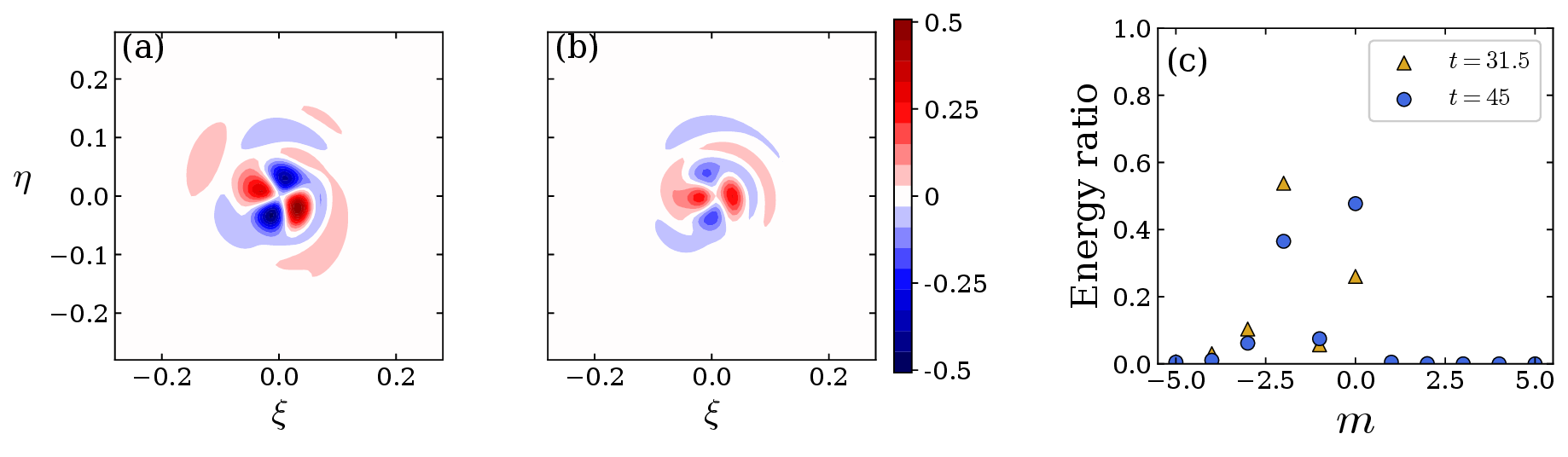}}
  \caption{Elliptic instability $(-2,0;[2,2])$ for $(\varepsilon,L/R)=(0.1,0.3)$. Contours of $\omega'_{B,2N}$ at (a)$t=31.5$, (b)$t=45$, and (b)energy ratios among the azimuthal components $\hat{u}_{B,2N}^{\prime(m)}$ in the local coordinates.}
\label{fig:WNL_elliptic_2N}
\end{figure}

For the two cases $(\varepsilon, L/R)=(0.1,0.3)$ and $(0.15,0.55)$, the nonlinear evolution of the elliptic instability$(-2,0;[2,2])$ shows a qualitatively similar behavior, as discussed in \S~\ref{sec:overview}.
In both cases, the growth of the unstable component saturates within the weakly nonlinear regime. 
Therefore, we focus on $(\varepsilon, L/R)=(0.1,0.3)$ as a representative case of the weakly nonlinear evolution of the elliptic instability.

Figure \ref{fig:WNL_elliptic_energy}(a) shows the time evolution of the disturbance energies $E'_{n}$ for $n=1$,$2$, and $3$.
The energy of the unstable component associated with the elliptic instability, $E'_1$, initially grows exponentially, at a rate consistent with the linear stability analysis.
The energies of the higher harmonic components, $E'_2$ and $E'_3$, also increase through nonlinear interactions involving the unstable component.
As $E'_1$ grows, however, its growth is gradually suppressed, eventually leading to saturation followed by decay.
The flow structure near saturation is visualized by iso-surfaces of $|\boldsymbol{\omega}|$ in figure \ref{fig:WNL_elliptic_energy}(b).
Although the helical vortex is slightly deformed by the disturbance, the deformation remains weak and no small-scale structures are observed.

To examine this process in more detail, figure~\ref{fig:WNL_elliptic_bfield} shows the vorticity component in the $\boldsymbol{e}_B$ direction for the total flow, $\omega_B$, and for the unstable component associated with the elliptic instability, $\omega'_{B,N}$, expressed in the local coordinates.
The vorticity distribution of the total flow field is slightly deformed as the disturbance grows, as seen at $t=25.5$ and $t=37.5$. 
After the disturbance energy begins to decay, the deformation becomes weaker, and the vorticity distribution at $t=49.5$ again resembles that observed in the linear regime at $t=13.5$.
For the unstable component, the vorticity distribution retains the characteristic structure of the elliptic instability $(-2,0;[2,2])$ at $t=13.5$ and $t=25.5$.
Even after the energy of the unstable component saturates, the vorticity distribution appears to remain close to this characteristic structure.

To characterize the mode structure quantitatively, the mode function $\tilde{\boldsymbol{u}}'_{N}$ is decomposed with respect to the local azimuthal coordinate $\psi$ as

\begin{equation}\label{eq:decomposition_local}
\tilde{\boldsymbol{u}}'_{N}(\rho,\psi,t) =\sum\limits_{m} \hat{\boldsymbol{u}}'^{(m)}_{N}(\rho,t) \exp{(\mathrm{i}m\psi)}.
\end{equation}
The structure of $\tilde{\boldsymbol{u}}'_{N}$ in the local coordinates is independent of the choice of point $P$, as discussed in \S~\ref{sec:base flow} and \S~\ref{sec:linear instability}.
Since the elliptic instability $(-2,0;[2,2])$ consists of Kelvin waves with azimuthal wavenumbers $m=-2$ and $m=0$, the unstable mode is expected to have the form $\tilde{\boldsymbol{u}}'_{N}\approx \hat{\boldsymbol{u}}'^{(-2)}_{N} \exp{(-2\mathrm{i}\psi)} + \hat{\boldsymbol{u}}'^{(0)}_{N}$ in the linear regime.
Indeed, figure~\ref{fig:WNL_elliptic_kelvin}(a) shows the energy ratio among the azimuthal components, confirming that the unstable mode is mainly composed of the $m=-2$ and $m=0$ components at $t=10.5$, in good agreement with the linear stability analysis.
Near saturation, at $t=31.5$ and $45$, the unstable mode remains dominated by these two components, with a slight change in their energy ratio from that in the linear regime.
Figures~\ref{fig:WNL_elliptic_kelvin}(b) and (c) show the radial distributions of $\hat{u}'^{(m)}_{B,N}$ for $m=-2$ and $m=0$, respectively.
In the linear regime, at $t=10.5$, these radial profiles are in good agreement with those obtained from the linear stability analysis.
Even near saturation, at $t=31.5$ and $45$, they are only slightly deformed from the linear-regime profiles, suggesting that the original mode structure is largely preserved.

\begin{figure}
  \centerline{\includegraphics[scale=.32]{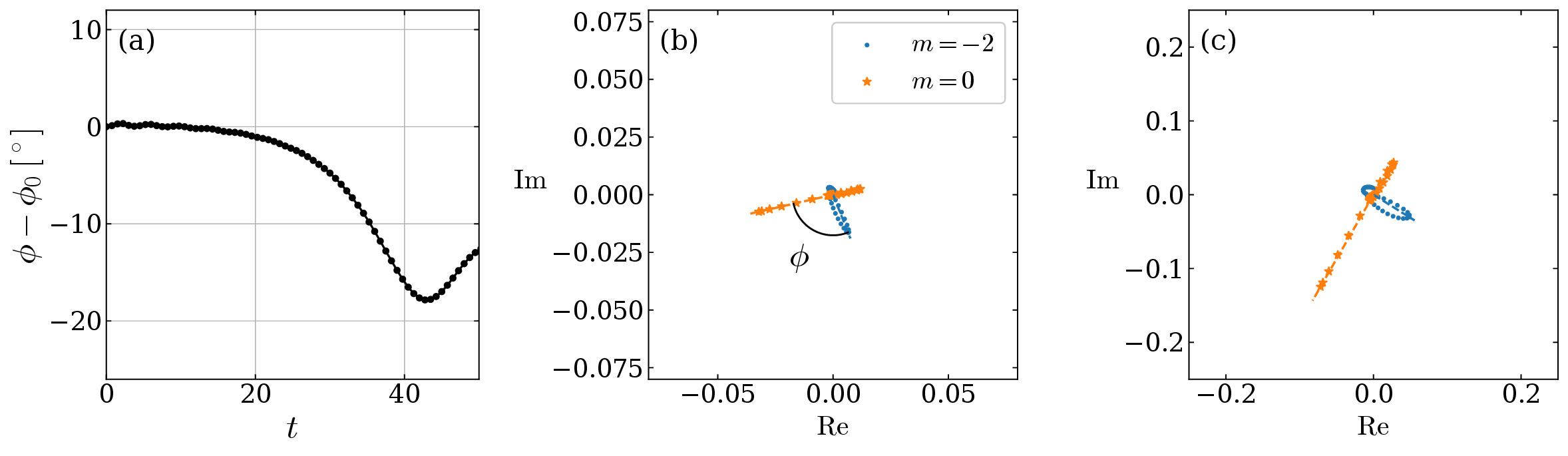}}
  \caption{Elliptic instability $(-2,0;[2,2])$ for $(\varepsilon,L/R)=(0.1,0.3)$. Phase difference between $\hat{u}'^{(-2)}_{B,N}$ and $\hat{u}'^{(0)}_{B,N}$: (a) time evolution of $\phi-\phi_0$, where $\phi_0$ is the initial value of $\phi$; complex-plane representation of $\hat{u}'^{(-2)}_{B,N}$ and $\hat{u}'^{(0)}_{B,N}$ at (b) $t=9$ and (c) $t=36$. The phase difference is defined between the points at which the magnitudes of the respective components attain their maxima.}
\label{fig:WNL_elliptic_phase}
\end{figure}

Figure \ref{fig:WNL_elliptic_2N}(a) and (b) show $\omega'_{B,2N}$ in the local coordinates at $t=31.5$ and $45$, respectively.
It can be seen that this higher harmonic component has a relatively simple spatial structure.
This observation is further supported by figure~\ref{fig:WNL_elliptic_2N}(c), which shows the energy ratios among the azimuthal components.
The energy is concentrated in only a small number of azimuthal components, indicating that the higher harmonic also retains a relatively simple mode structure.
Together with the preserved structure of the unstable component, this suggests that the nonlinear evolution is dominated by interactions among a limited number of modes, as expected in the weakly nonlinear regime.

The above results show that the saturation of the disturbance energy occurs through a weakly nonlinear process.
Such saturation may be associated with phase detuning between the resonant Kelvin waves induced by weakly nonlinear effects \citep{sipp_weakly_2000}.
To examine this possibility, figure \ref{fig:WNL_elliptic_phase} shows the time evolution of the phase difference $\phi$ between $\hat{u}'^{(-2)}_{B,N}$ and $\hat{u}'^{(0)}_{B,N}$.
During the linear regime, the phase difference remains nearly constant because the two components oscillate at the same frequency.
However, the phase difference starts to vary as the disturbance grows and changes appreciably around the time at which the disturbance energy saturates.
The coincidence between the phase variation and the saturation of the disturbance energy suggests that phase detuning between the Kelvin waves is involved in the saturation process.
In physical terms, the change in their relative phase rotates the disturbance structure away from the orientation of the linear eigenmode, consistent with the weakly nonlinear mechanism described by \citet{sipp_weakly_2000, schaeffer_nonlinear_2010}.

\subsubsection{Curvature instability}\label{sec:WNL_curvature}
For the curvature instability $(-1,0;[2,4])$, the two cases $(\varepsilon, L/R)=(0.11,0.5)$ and $(0.13,0.5)$ exhibit a qualitatively similar weakly nonlinear evolution, as discussed in \S~\ref{sec:overview}.
We therefore focus on $(\varepsilon, L/R)=(0.11,0.5)$ as a representative case.

\begin{figure}
  \centerline{\includegraphics[scale=.45]{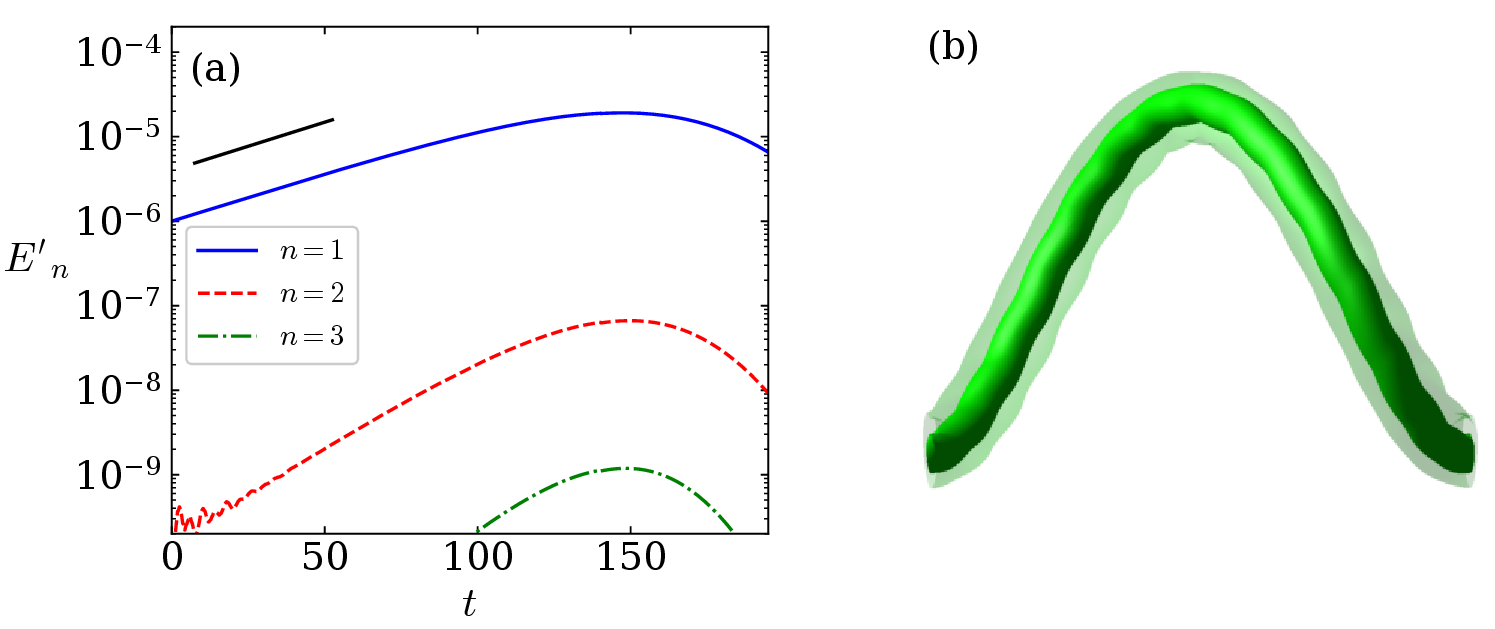}}
  \caption{Curvature instability $(-1,0;[2,4])$ for $(\varepsilon, L/R)=(0.11,0.5)$. (a)Time evolution of the disturbance energies $E'_n$.
  The black line indicates the slope corresponding to the energy growth predicted by the linear stability analysis.
  (b)Isosurfaces of the vorticity magnitude $|\boldsymbol{\omega}|$ at $t=150$, when the energy of the unstable component is close to its maximum. The translucent and opaque isosurfaces correspond  to$|\boldsymbol{\omega}|/|\boldsymbol{\omega}_0|_{\max}=0.1$ and $|\boldsymbol{\omega}|/|\boldsymbol{\omega}_0|_{\max}=0.5$, respectively.}
\label{fig:WNL_curvature_energy}
\end{figure}

\begin{figure}
  \centerline{\includegraphics[scale=.28]{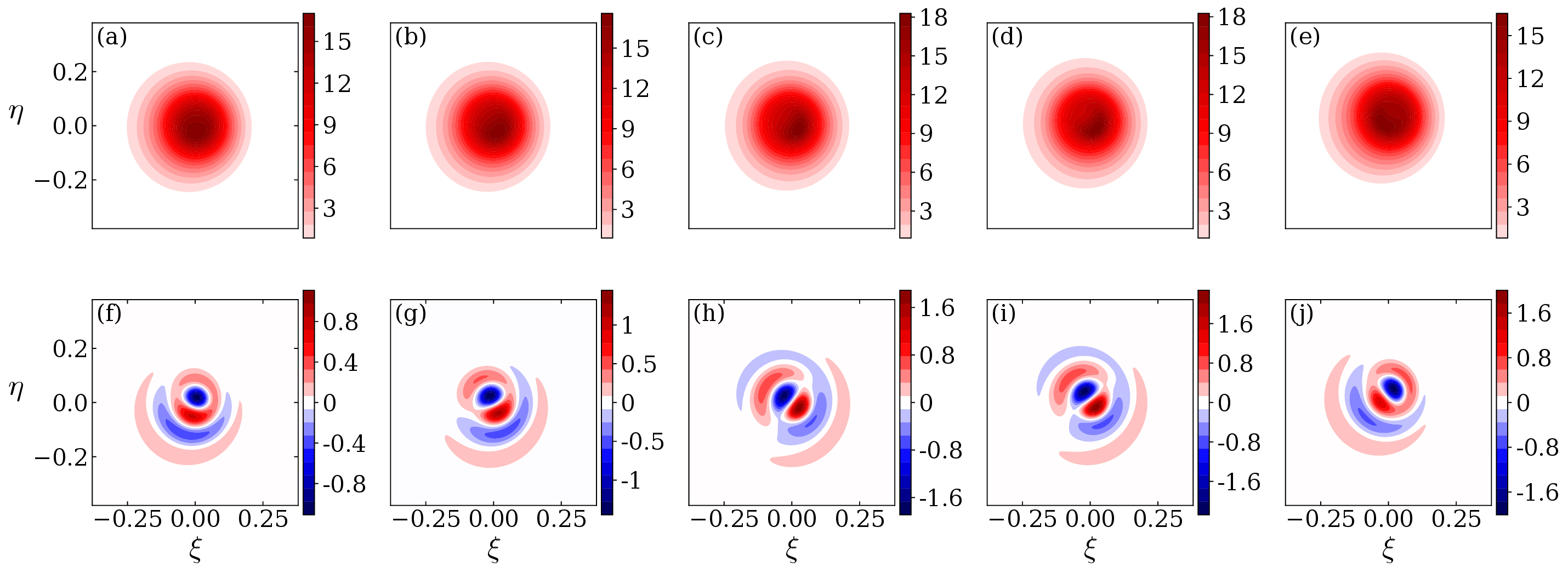}}
  \caption{Curvature instability $(-1,0;[2,4])$ for $(\varepsilon, L/R)=(0.11,0.5)$. Contours of (a--e) the total vorticity $\omega_B$ and (f--j) the disturbance vorticity $\omega'_{B,N}$ in the local coordinates.
      From left to right: $t=37.5,75,112.5,150,187.5$.}
\label{fig:WNL_curvature_bfield}
\end{figure}

\begin{figure}
  \centerline{\includegraphics[scale=.39]{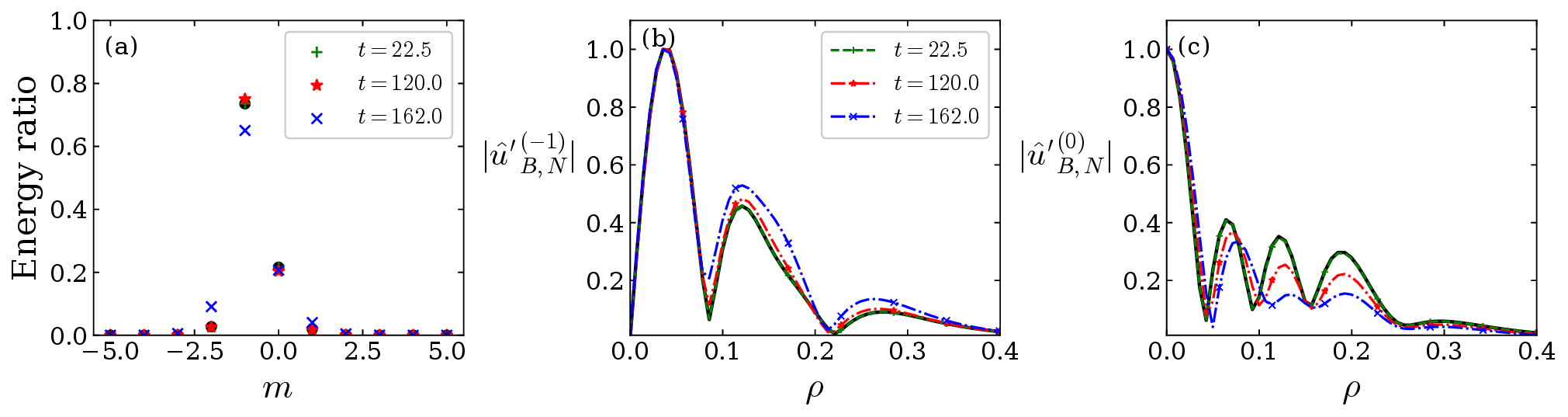}}
  \caption{Curvature instability $(-1,0;[2,4])$ for $(\varepsilon, L/R)=(0.11,0.5)$. Structure of the unstable mode in the local coordinates: (a)energy ratios among the azimuthal components $\hat{u}_{B,N}^{\prime(m)}$ and radial distribution of $|\hat{u}_{B,N}^{\prime(m)}|$ for (b) $m=-1$ and (c) $m=0$. Black circles and solid lines denote the results of the linear stability analysis.}
\label{fig:WNL_curvature_kelvin}
\end{figure}

\begin{figure}
  \centerline{\includegraphics[scale=.32]{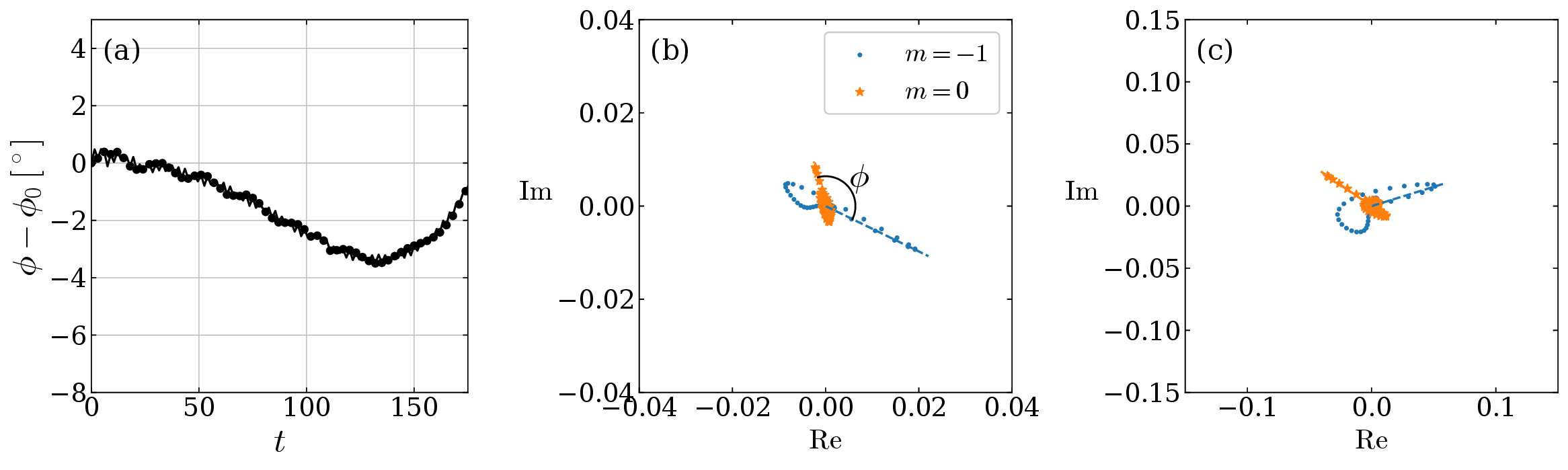}}
  \caption{Curvature instability $(-1,0;[2,4])$ for $(\varepsilon, L/R)=(0.11,0.5)$. Phase difference between $\hat{u}'^{(-1)}_{B,N}$ and $\hat{u}'^{(0)}_{B,N}$: (a) time evolution of $\phi-\phi_0$; complex-plane representation of $\hat{u}'^{(-1)}_{B,N}$ and $\hat{u}'^{(0)}_{B,N}$ at (b) $t=15$ and (c) $t=165$. The phase difference is defined between the points at which the magnitudes of the respective components attain their maxima.}
\label{fig:WNL_curvature_phase}
\end{figure}

Overall, despite the different modal compositions of the two instabilities, the curvature instability exhibits the same qualitative characteristics of weakly nonlinear saturation as the elliptic instability discussed in \S~\ref{sec:WNL_elliptic}.
As shown in figure~\ref{fig:WNL_curvature_energy}(a), the unstable component $(n=1)$ associated with the curvature instability grows at a rate consistent with the linear stability analysis, while its harmonic components are generated through nonlinear interactions, before its growth is suppressed and the disturbance energy subsequently decays.
Throughout this evolution, the helical vortex is only slightly deformed and the vorticity distribution of the unstable component appears to remain close to the characteristic structure of the curvature instability, as shown in figures \ref{fig:WNL_curvature_energy}(b) and \ref{fig:WNL_curvature_bfield}.

Using the decomposition introduced in equation~\eqref{eq:decomposition_local}, the structure of the unstable mode is examined in figure \ref{fig:WNL_curvature_kelvin}.
Since the curvature instability $(-1,0;[2,4])$ consists of Kelvin waves with azimuthal wavenumbers $m=-1$ and $m=0$, the unstable mode in the linear regime is expected to have the form $\tilde{\boldsymbol{u}}'_{N}\approx \hat{\boldsymbol{u}}'^{(-1)}_{N} \exp{(-\mathrm{i}\psi)} + \hat{\boldsymbol{u}}'^{(0)}_{N}$.
Indeed, figure \ref{fig:WNL_curvature_kelvin}(a) shows that the unstable mode is dominated by the $m=-1$ and $m=0$ components in the linear regime, as seen at $t=22.5$, in agreement with the linear stability analysis.
Near saturation, at $t=120$ and $162$, these two components remain dominant, with only a slight change in their energy ratio.
Figures~\ref{fig:WNL_curvature_kelvin}(b) and (c) show the radial profiles of these two components.
The profiles near saturation are only slightly modified from those in the linear regime, indicating that the characteristic structure of the unstable mode is largely preserved.
Together with the hierarchical growth of the higher harmonic energies in figure~\ref{fig:WNL_curvature_energy}(a), these results indicate that the nonlinear evolution remains dominated by a limited set of modes. 
This provides the same characteristic signature of weakly nonlinear evolution as observed for the elliptic instability.

To examine whether a similar phase-detunig process accompanies the saturation of the curvature instability, figure~\ref{fig:WNL_curvature_phase} shows the time evolution of the phase difference between the interacting Kelvin wave components. 
As in the elliptic instability, the phase difference remains nearly constant during the linear regime but begins to vary as the disturbance approaches saturation.
This behavior is consistent with that observed for the elliptic instability, further indicating that the two instabilities undergo qualitatively similar weakly nonlinear saturation.

\subsection{Beyond the weakly nonlinear regime}
For both the elliptic and curvature instabilities, some cases exhibit nonlinear evolution that proceeds beyond the weakly nonlinear regime, resulting in significant deformation of the original vortex structure.
These cases are generally associated with relatively large linear growth rates.
In this section, we examine their dynamics in detail and show how their nonlinear evolution differs qualitatively from that observed within the weakly nonlinear regime.

\subsubsection{Elliptic instability}\label{sec:BWNL_elliptic}

\begin{figure}
  \centerline{\includegraphics[scale=.45]{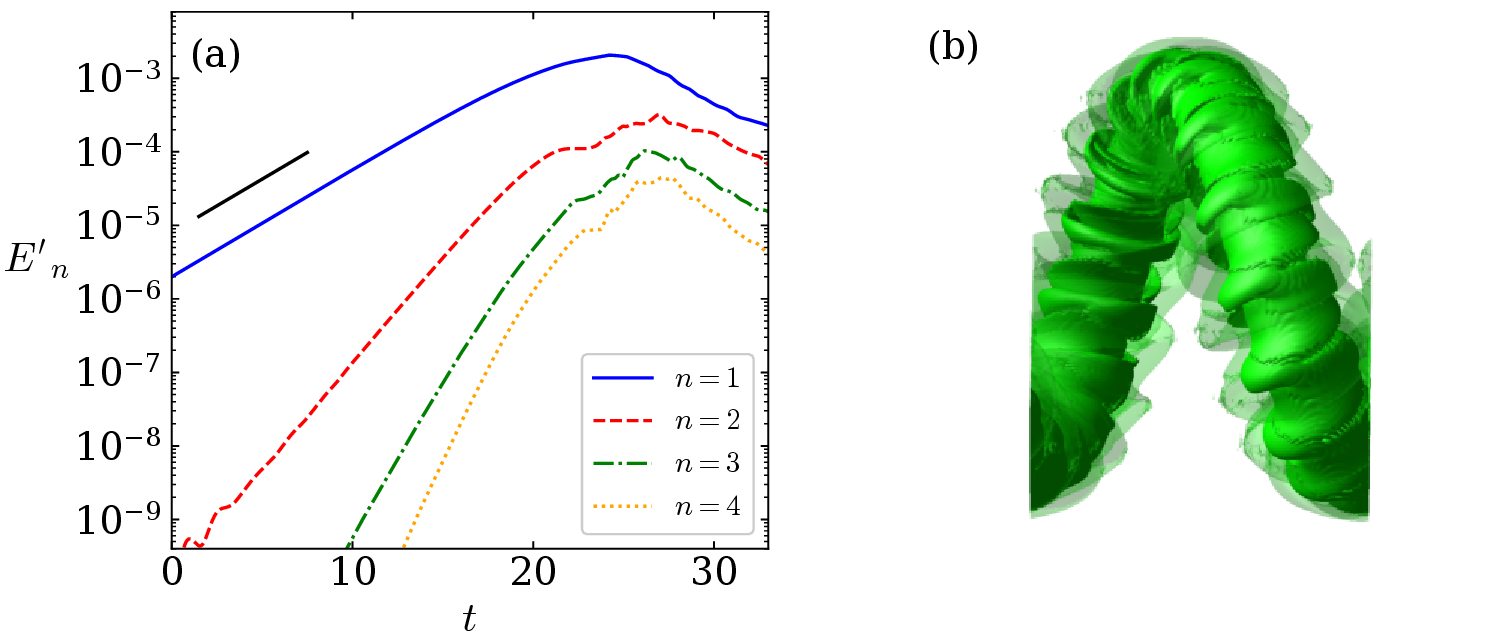}}
  \caption{Elliptic instability $(-2,0;[2,2])$ for $(\varepsilon,L/R)=(0.2,0.3)$. (a)Time evolution of the disturbance energies $E'_n$.
  The black line indicates the slope corresponding to the energy growth predicted by the linear stability analysis.
  (b)Isosurfaces of the vorticity magnitude $|\boldsymbol{\omega}|$ at $t=24.75$, when the energy of the unstable component is close to its maximum. The translucent and opaque isosurfaces correspond  to$|\boldsymbol{\omega}|/|\boldsymbol{\omega}_0|_{\max}=0.15$ and $|\boldsymbol{\omega}|/|\boldsymbol{\omega}_0|_{\max}=0.5$, respectively.}
\label{fig:BWNL_elliptic_energy}
\end{figure}

For the three cases $(\varepsilon,L/R)=(0.15,0.3)$, $(0.2,0.3)$ and $(0.15,0.2)$, the nonlinear evolution of the elliptic instability is accompanied by an increase in enstrophy and a greater contribution from higher harmonic components, as discussed in \S~\ref{sec:overview}.
These features indicate that the evolution proceeds beyond the weakly nonlinear regime.
We therefore focus on $(\varepsilon, L/R)=(0.2, 0.3)$ as a representative case in which nonlinear effects are relatively significant.

Figure \ref{fig:BWNL_elliptic_energy}(a) shows the time evolution of $E'_n$ for $n=1,2,3$ and $4$.
As in the weakly nonlinear case discussed in \S~\ref{sec:WNL_elliptic}, $E'_1$ initially grows exponentially at a rate consistent with the linear stability analysis, while the higher harmonic components are generated through nonlinear interactions.
A significant difference is that $E'_1$ attains a substantially higher level, accompanied by appreciable growth of the higher harmonic components.
The corresponding flow structure is visualized by isosurfaces of $|\boldsymbol{\omega}|$ at $t=24.75$ in figure \ref{fig:BWNL_elliptic_energy}(b).
At this stage, the helical vortex is significantly deformed, in contrast to the evolution within the weakly nonlinaer regime.

\begin{figure}
  \centerline{\includegraphics[scale=.25]{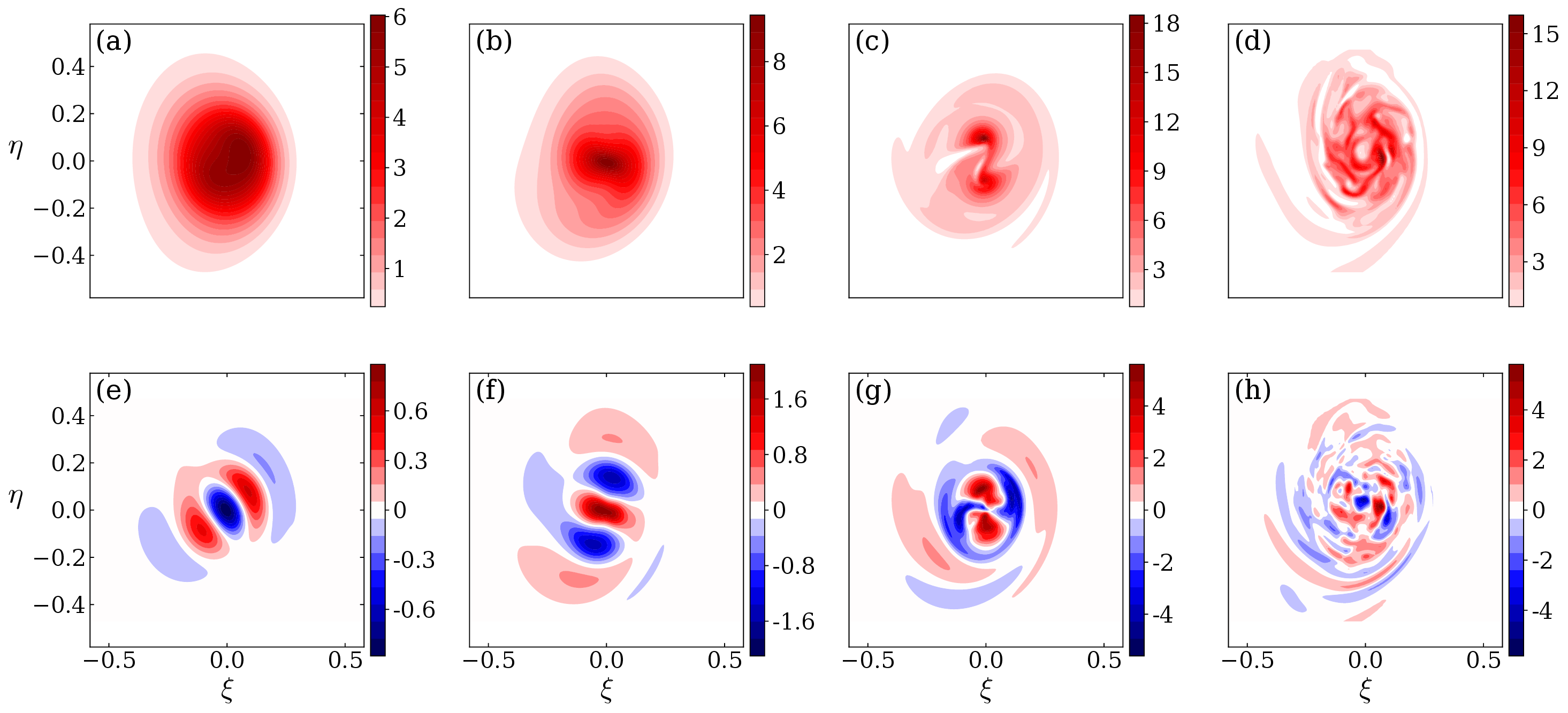}}
  \caption{Elliptic instability $(-2,0;[2,2])$ for $(\varepsilon,L/R)=(0.2,0.3)$. Contours of (a--d) the total vorticity $\omega_B$ and (e--h) the disturbance vorticity $\omega'_{B,N}$ in the local coordinates.
  From left to right: $t=7.5,15,22.5,30$.}
\label{fig:BWNL_elliptic_bfield}
\end{figure}

\begin{figure}
  \centerline{\includegraphics[scale=.33]{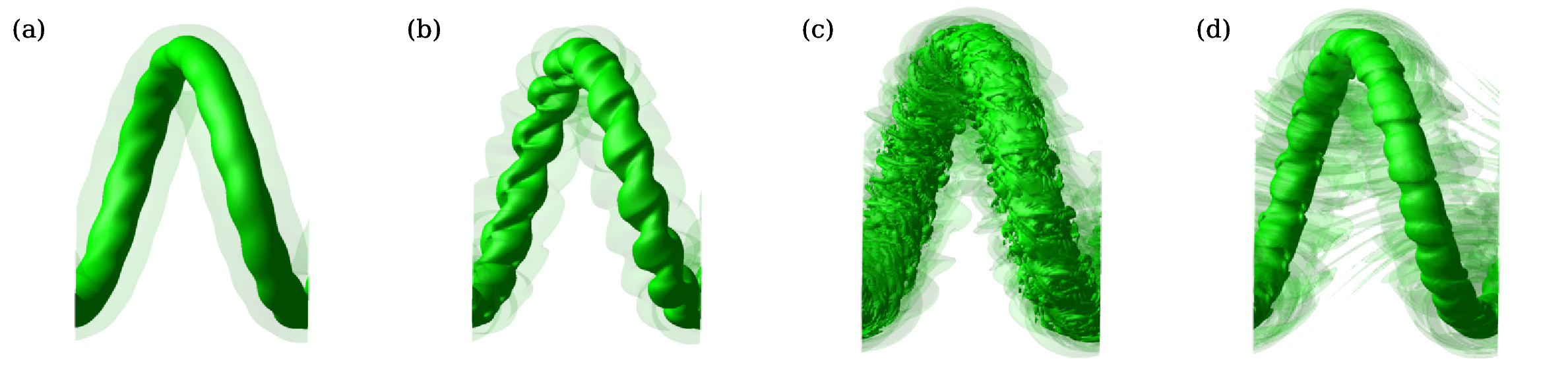}}
  \caption{Elliptic instability $(-2,0;[2,2])$ for $(\varepsilon,L/R)=(0.2,0.3)$. Isosurfaces of the vorticity magnitude $|\boldsymbol{\omega}|$. The translucent and opaque isosurfaces correspond  to$|\boldsymbol{\omega}|/|\boldsymbol{\omega}_0|_{\max}=0.15$ and $|\boldsymbol{\omega}|/|\boldsymbol{\omega}_0|_{\max}=0.6$, respectively. From left to right:$t=6, 17.25, 30, 45$.}
\label{fig:BWNL_elliptic_iso}
\end{figure}

To examine how the disturbance growth modifies the vortex core, figure \ref{fig:BWNL_elliptic_bfield} shows the temporal evolution of $\omega_B$ and $\omega'_{B,N}$ in the local coordinates.
The vorticity distribution of the total flow field is gradually deformed as the disturbance grows.
Around $t=22.5$, it departs substantially from the initial nearly axisymmetric distribution, with the vorticity becoming concentrated in regions associated with the disturbance.
The deformation subsequently becomes increasingly complex, and small-scale structures eventually emerge as the vortex core undergoes severe deformation at $t=30$.
For the unstable component, the vorticity distribution remains close to the characteristic structure of the elliptic instability $(-2,0;[2,2])$ at $t=7.5$ and $t=15$. 
Its deformation becomes apparent around $t=22.5$, and the organized disturbance structure eventually breaks down into small-scale structures at $t=30$.

The corresponding three-dimensional evolution is shown in figure~\ref{fig:BWNL_elliptic_iso}.
In the linear regime, the helical vortex is only slightly deformed, as seen at $t=6$.
As the disturbance grows, a double-helix structure characteristic of the elliptic instability $(-2,0;[2,2])$ develops around $t=17.25$.
The deformation subsequently becomes more complex, and small-scale structures are eventually generated around $t=30$.
After the formation of these small-scale structures, the vortex undergoes relaminarization, as seen at $t=45$.
This suggests that an instability-breakdown-relaminarization process \citep{leweke_dynamics_2016} can also occur in the presence of axial flow.

\subsubsection{Curvature instability}\label{sec:BWNL_curvature}

For the curvature instability, the case $(\varepsilon,L/R)=(0.16,0.50)$ exhibits nonlinear evolution beyond the weakly nonlinear regime, as discussed in \S~\ref{sec:overview}. 
We therefore focus on this case below.

\begin{figure}
  \centerline{\includegraphics[scale=.45]{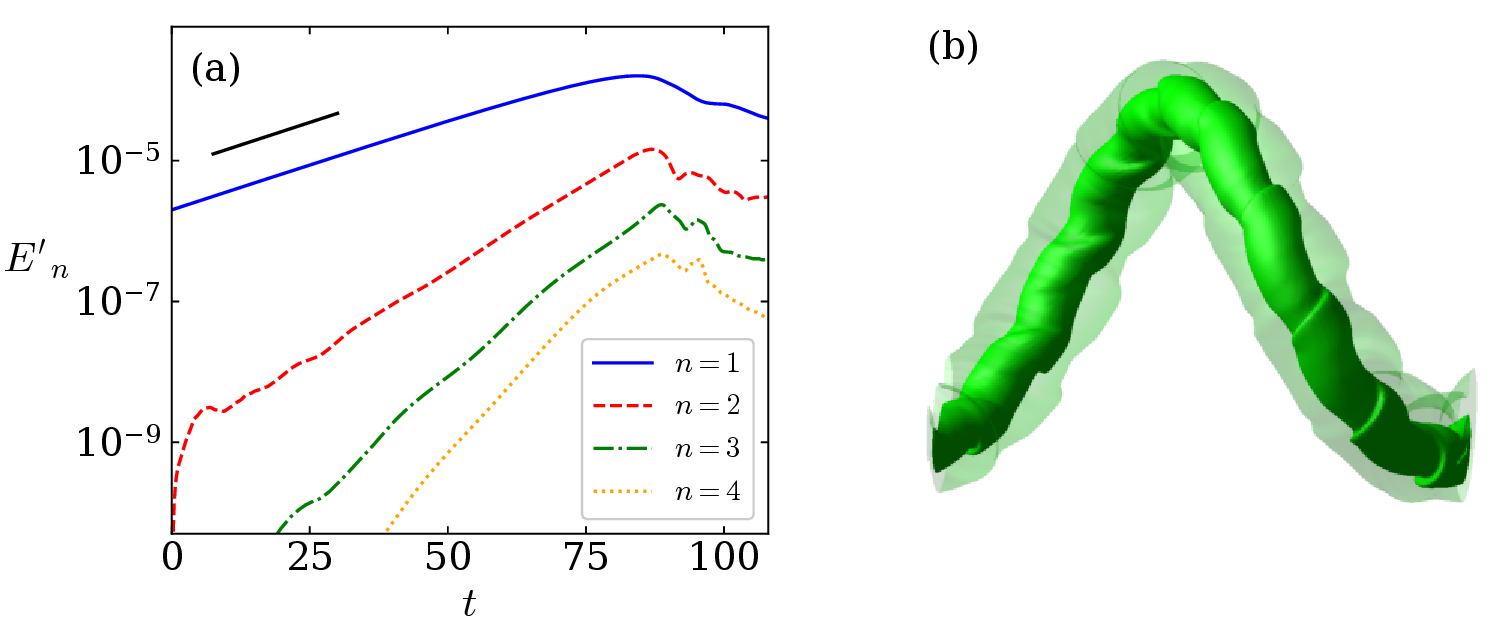}}
  \caption{Curvature instability $(-1,0;[2,4])$ for $(\varepsilon,L/R)=(0.16,0.5)$. (a)Time evolution of the disturbance energies $E'_n$.
  The black line indicates the slope corresponding to the energy growth predicted by the linear stability analysis.
  (b)Isosurfaces of the vorticity magnitude $|\boldsymbol{\omega}|$ at $t=82.5$, when the energy of the unstable component is close to its maximum. The translucent and opaque isosurfaces correspond  to$|\boldsymbol{\omega}|/|\boldsymbol{\omega}_0|_{\max}=0.1$ and $|\boldsymbol{\omega}|/|\boldsymbol{\omega}_0|_{\max}=0.5$, respectively.}
\label{fig:BWNL_curvature_energy}
\end{figure}

\begin{figure}
  \centerline{\includegraphics[scale=.28]{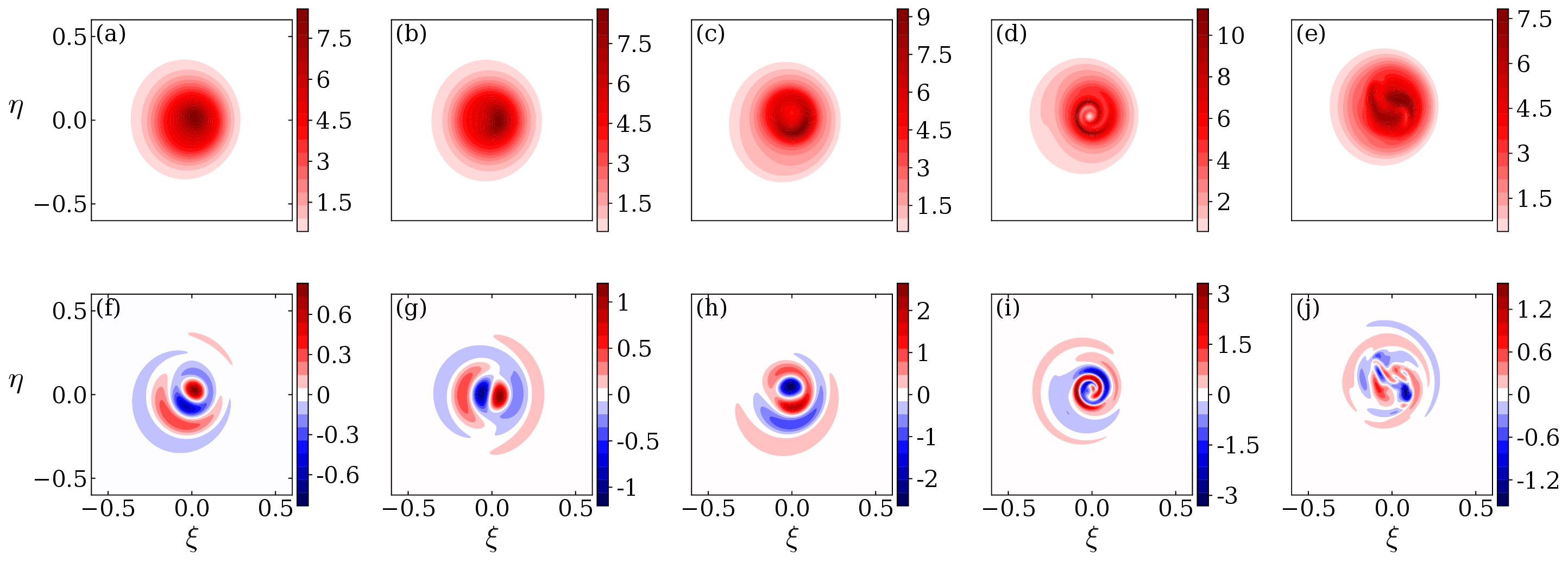}}
  \caption{Curvature instability $(-1,0;[2,4])$ for $(\varepsilon,L/R)=(0.16,0.5)$. Contours of (a--e) the total vorticity $\omega_B$ and (f--j) the disturbance vorticity $\omega'_{B,N}$ in the local coordinates.
  From left to right: $t=25.5,48,70.5,93,115.5$.}
\label{fig:BWNL_curvature_bfield}
\end{figure}

As in the elliptic instability case discussed in \S~\ref{sec:BWNL_elliptic}, the initial evolution is qualitatively similar to that in the weakly nonlinear regime.
As shown in figure \ref{fig:BWNL_curvature_energy}(a), $E'_1$ initially grows at the linear growth rate, while higher harmonic components are generated through nonlinear interactions.
As the disturbance grows, however, $E'_1$ reaches a larger level than in the weakly nonlinear cases, accompanied by appreciable growth of the higher harmonic components.
The corresponding flow structure near the maximum of $E'_1$ is shown in \ref{fig:BWNL_curvature_energy}(b), where the helical vortex is appreciably deformed.

The time evolutions of $\omega_B$ and $\omega'_{B,N}$ in local coordinates are shown in figure~\ref{fig:BWNL_curvature_bfield}.
During the linear regime, at $t=25.5$ and $48$, the total vorticity is only weakly deformed, while $\omega'_{B,N}$ retains the characteristic structure of the curvature instability.
As the disturbance grows, the deformation of the total vorticity becomes increasingly pronounced, while the characteristic structure of the unstable component becomes progressively distorted and is no longer preserved near the maximum of $E'_1$, as seen at $t=93$ and $115.5$.
This behavior differs qualitatively from that in the weakly nonlinear case, where the unstable mode structure is largely unchanged near saturation.
Although generation of small-scale structures is less evident than in the elliptic instability case, the loss of the characteristic modal structure, together with the appreciable growth of the higher harmonic components, indicates that the evolution has proceeded beyond the weakly nonlinear regime.
This point is further discussed in the next section.

\begin{figure}
  \centerline{\includegraphics[scale=.33]{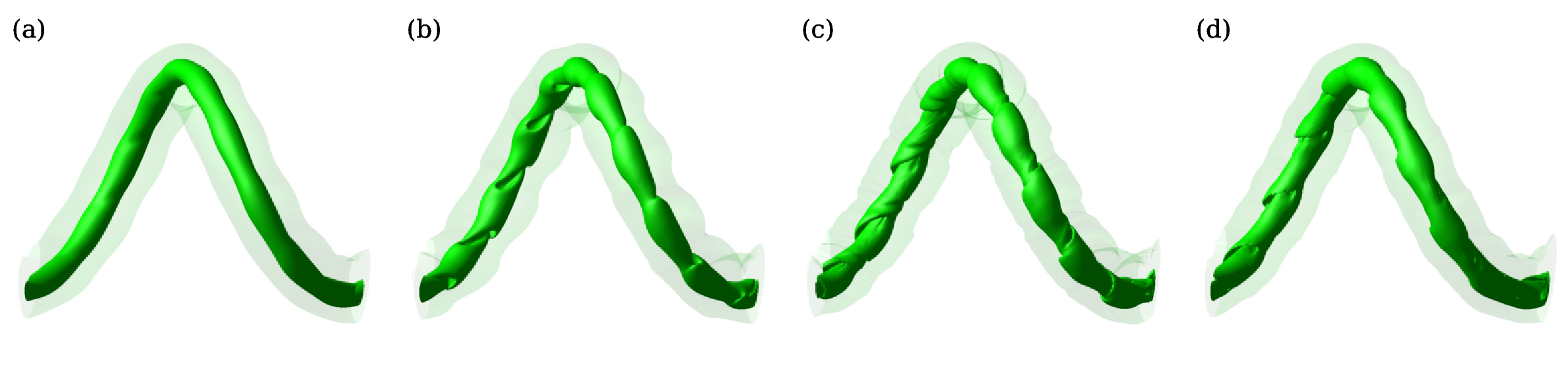}}
  \caption{Curvature instability $(-1,0;[2,4])$ for $(\varepsilon,L/R)=(0.16,0.5)$. Isosurfaces of the vorticity magnitude $|\boldsymbol{\omega}|$. The translucent and opaque isosurfaces correspond  to$|\boldsymbol{\omega}|/|\boldsymbol{\omega}_0|_{\max}=0.1$ and $|\boldsymbol{\omega}|/|\boldsymbol{\omega}_0|_{\max}=0.7$, respectively. From left to right:$t=37.5,71.25,90,108.75$.}
\label{fig:BWNL_curvature_iso}
\end{figure}

The corresponding three-dimensional evolution is shown by the iso-surfaces in figure~\ref{fig:BWNL_curvature_iso}.
A slight deformation of the helical vortex is visible during the linear regime.
As the disturbance grows, the characteristic helically deformed structure of the curvature instability \citep{hattori_numerical_2019} becomes apparent at $t=71.25$.
The vortex subsequently undergoes substantially stronger deformation than in the weakly nonlinear case, before the deformation becomes weaker as the disturbance decays.

\section{Discussion}\label{sec:discussion}
The results presented above show weakly nonlinear dynamics in several cases.
We now compare these numerical results with predictions from weakly nonlinear analysis.
The weakly nonlinear evolution of short-wave instabilities can be described by amplitude equations governing the resonant Kelvin waves and the associated mean-flow correction.
Such a formulation was developed for the elliptic instability of a strained Lamb-Oseen vortex by \citet{sipp_weakly_2000}.
More recently, the same framework was applied to the curvature instability, and the predicted weakly nonlinear behavior was shown to be consistent with numerical results \citep{hattori_weakly_2026}. 
Although the coefficients depend on the base flow and the instability mode, the amplitude equations obtained in these studies share the same structure, with damping terms included:

\begin{subequations}\label{eq:amplitude_equations}
\begin{align}
\frac{\mathrm{d}A}{\mathrm{d}t}
&=
-\mu_A A
+ \mathrm{i} \kappa_{A} B
+
\left(
\mathrm{i}c_{A1}|A|^2
+
\mathrm{i}c_{A2}|B|^2
+
\mathrm{i}c_{A3}C
\right)A,
\label{eq:amp_A}
\\
\frac{\mathrm{d}B}{\mathrm{d}t}
&=
-\mu_B B
+ \mathrm{i} \kappa_{B} A
+
\left(
\mathrm{i}c_{B1}|A|^2
+
\mathrm{i}c_{B2}|B|^2
+
\mathrm{i}c_{B3}C
\right)B,
\label{eq:amp_B}
\\
\frac{\mathrm{d}C}{\mathrm{d}t}
&=
\mathrm{i}
c_{C}
\left(
A\overline{B}
-
\overline{A}B
\right).
\label{eq:amp_C}
\end{align}
\end{subequations}
Here, $A$ and $B$ denote the complex amplitudes of the resonant Kelvin waves, and $C$ represents the mean-flow correction generated by their nonlinear interaction.
The coefficients $\mu_A$ and $\mu_B$ account for linear damping, while $\kappa_{A}$ and $\kappa_{B}$ characterize the linear coupling between $A$ and $B$.
The coefficients $c_{Ai}$, $c_{Bi}$ and $c_C$ characterize the nonlinear interactions.
A brief summary of the weakly nonlinear formulation relevant to the present comparison is given in appendix~\ref{appC}.
In the following, we apply these equations to the curvature instability mode $(-1,0;[2,4])$.

\begin{figure}
  \centerline{\includegraphics[scale=.4]{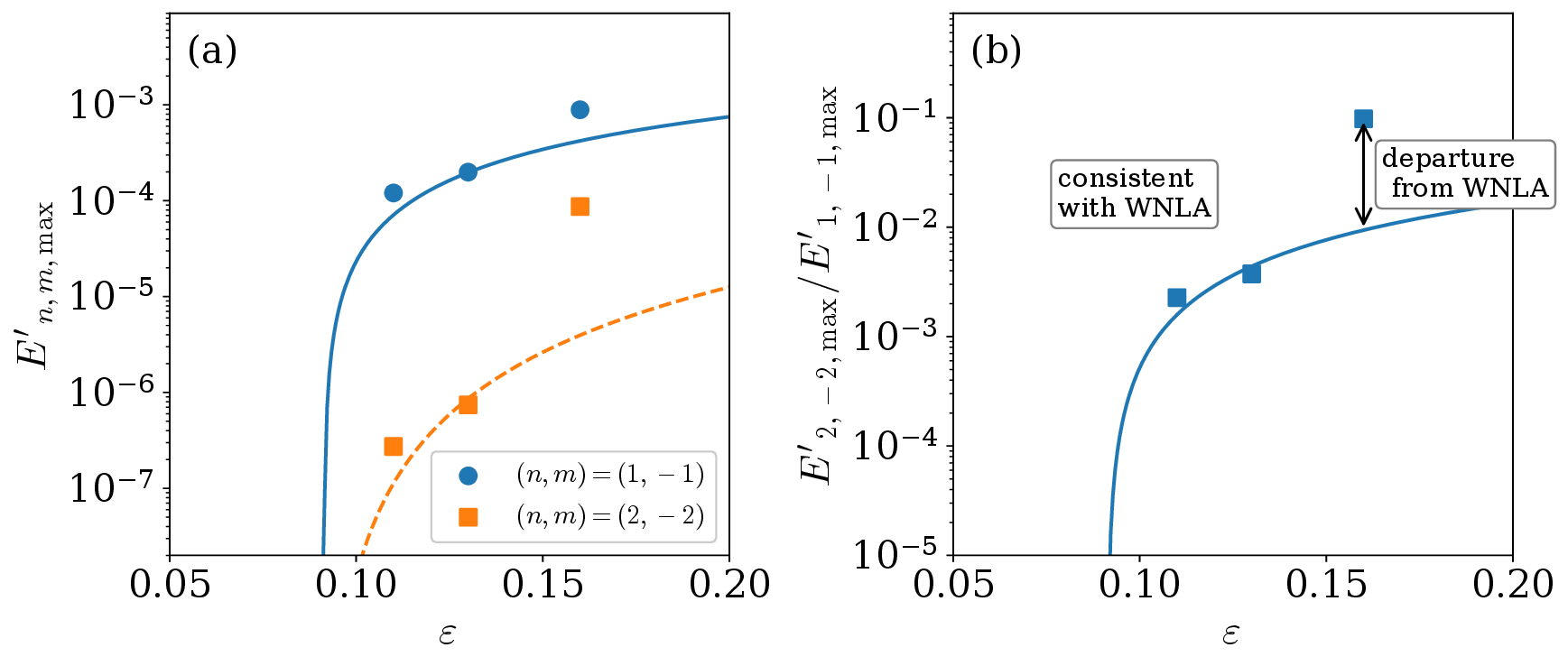}}
  \caption{Comparison of numerical results and weakly nonlinear analysis (WNLA) for the curvature instability.
(a) Maximum energies $E'_{n,m,\max}$ plotted against $\varepsilon$.
Symbols denote the numerical results, while solid and dashed curves denote the corresponding WNLA predictions.
Both the numerical and WNLA quantities are scaled using the vortex-core scales.
(b) Ratio $E'_{2,-2,\max}/E'_{1,-1,\max}$ plotted against $\varepsilon$; squares denote the numerical results, while the solid curve denotes the WNLA prediction.}
\label{fig:VS_WNL}
\end{figure}

For comparison with the weakly nonlinear analysis, the disturbance energy of the local azimuthal component is introduced as

\begin{equation}\label{eq:decomposition_local_ene}
E'_{n,m} = \int |\hat{\boldsymbol{u}}'^{(m)}_{nN}(\rho,t)|^2 \rho\,\mathrm{d}\rho. 
\end{equation}
Here, the quantities are nondimensionalized using the vortex-core scales \citep{hattori_numerical_2019}, as in the weakly nonlinear analysis.
The maximum energies of the unstable component, $E'_{1,-1,\max}$, and its higher harmonic, $E'_{2,-2,\max}$, are compared with the weakly nonlinear predictions in figure~\ref{fig:VS_WNL}.
The weakly nonlinear analysis predicts that $E'_{1,-1,\max}$ increases with $\varepsilon$, and this trend is consistent with the numerical results, as shown in figure~\ref{fig:VS_WNL}(a).
A similar increase is predicted and observed for the higher-harmonic component $E'_{2,-2,\max}$.
However, the quantitative agreement for $E'_{2,-2, \max}$ deteriorates at $\varepsilon=0.16$.
This discrepancy is more clearly seen in figure~\ref{fig:VS_WNL}(b), which shows the ratio $E'_{2,-2,\max}/E'_{1,-1,\max}$.
At $\varepsilon=0.16$, a substantially larger fraction of the disturbance energy is contained in the higher-harmonic component in the numerical results than predicted by the weakly nonlinear analysis.
This provides quantitative evidence that higher harmonic components participate more strongly in the nonlinear evolution than considered in the weakly nonlinear description, consistent with the classification of this case as being beyond the weakly nonlinear regime.

This discrepancy may be attributed to the increasing importance of higher-order nonlinear effects as the disturbance amplitude becomes larger.
As introduced in appendix \ref{appC}, the amplitude equations~\eqref{eq:amplitude_equations} retain contributions only up to a given order in the asymptotic expansion.
The neglected terms are of higher order in the disturbance amplitudes $A$ and $B$ \citep{knobloch_normal_1994}, and their relative importance therefore increases as the amplitude gets larger.
For example, a larger $E'_{1,-1,\max}$ corresponds to a larger $|A|^2$, indicating the increasing importance of higher-order contributions.
Accounting for these higher-order effects would require carrying the asymptotic expansion to higher orders, thereby introducing additional nonlinear interactions and modal components.
This interpretation is consistent with the numerical results: as $E'_{1,-1,\max}$ increases, the energies of the higher harmonic components also increase as discussed in \S~\ref{sec:overview}, and the fourth harmonic component ($n=4$), which is not included in the present weakly nonlinear analysis, becomes appreciable.

Regarding the onset of the evolution beyond the weakly nonlinear regime, the linear growth rate appears to provide an important indicator, as discussed in \S~\ref{sec:overview} and also observed by \citet{schaeffer_nonlinear_2010}.
This tendency may be related to the balance between linear and nonlinear effects in the amplitude equations.
In the weakly nonlinear analysis, when the disturbance amplitude ceases to grow, the linear growth must be balanced by nonlinear effects; a larger linear growth rate therefore generally requires a larger disturbance amplitude for nonlinear effects to suppress the linear growth \citep{sipp_weakly_2000, hattori_weakly_2026}.
Even beyond the formal range of validity of the present amplitude equations, it is plausible that the same competition between linear and nonlinear effects persists, with higher-order terms also contributing to the nonlinear effects.
At such larger amplitudes, departure from the weakly nonlinear regime becomes more likely, as already discussed.
Therefore, parameter changes that increase the linear growth rate tend to lead to larger disturbance amplitudes and stronger nonlinear effects.

The same argument is expected to apply to the elliptic instability $(-2,0;[2,2])$, provided that critical-layer effects do not become dominant, since its amplitude equations have the same structure following the procedure in appendix \ref{appC}. A detailed derivation and quantitative comparison require separate investigation and are left for future work.

Overall, the above discussion provides a new physical interpretation of the departure from the weakly nonlinear regime to beyond it for short-wave instabilities, based primarily on the general structure of the amplitude equations.

\section{Concluding remarks}\label{sec:conclusion}
In this paper, we have studied nonlinear evolution of a helical vortex disturbed by short-wave instabilities.
A sufficiently small-amplitude elliptic instability mode $(-2,0;[2,2])$ or curvature instability mode $(-1,0;[2,4])$ is imposed as the initial disturbance, and the normalized vortex core radius $\varepsilon$ and reduced pitch $L/R$ are systematically varied to examine cases with a wide range of different linear growth rates.
As a result, two types of nonlinear evolution are observed for both the elliptic and curvature instabilities: evolution within the weakly nonlinear regime for relatively small linear growth rates and evolution beyond this regime for relatively large growth rates.

For the evolution within the weakly nonlinear regime, the overall structure of the helical vortex is largely preserved.
The short-wave instability mode initially grows linearly and subsequently saturates at a relatively low energy level, accompanied by progressively smaller energies in the higher harmonic components.
The structure of the short-wave instability mode is also largely preserved throughout the evolution, while nonlinear detuning between the resonant Kelvin waves is observed.
These results are consistent with previous results for the elliptic instability $(-1,1)$ \citep{schaeffer_nonlinear_2010,sipp_weakly_2000}, suggesting that the weakly nonlinear dynamics leading to saturation is a generic feature of short-wave instabilities.
Therefore, even in helical vortices with axial flow, as commonly found in practical applications, short-wave instabilities may saturate under certain conditions without leading to a turbulent transition.

For the evolution beyond the weakly nonlinear regime, the overall structure of the helical vortex is noticeably deformed, with an increase in enstrophy.
The short-wave instability mode initially grows linearly and attains larger energy level, accompanied by larger energies in the higher harmonic components.
The structure of the short-wave instability mode is significantly deformed through the evolution and eventually breaks down.
For the elliptic instability, small-scale structures are subsequently generated in the total flow, followed by an instability-breakdown-relaminarization process \citep{leweke_dynamics_2016}.

The observed nonlinear evolutions are further discussed through comparison with the weakly nonlinear analysis.
For the curvature instability, the maximum energies of the unstable component and its higher harmonic increase with $\varepsilon$ , corresponding to an increase in the linear growth rate, in both the present simulation and the weakly nonlinear analysis.
As the disturbance amplitude becomes larger, however, the numerical results increasingly depart from the weakly nonlinear predictions.
Within the weakly nonlinear framework, this departure suggests the increasing importance of higher-order nonlinear effects and additional modal interactions.
This is consistent with the numerical results, in which the fourth harmonic component ($n=4$), not retained in the present weakly nonlinear formulation, becomes increasingly appreciable.
In addition, the observed relation between the linear growth rate and the resulting nonlinear effects can be qualitatively understood from the structure of the amplitude equations: a larger linear growth rate tends to require a larger disturbance amplitude for nonlinear effects to suppress the growth and is therefore associated with stronger nonlinear evolution. 
The above discussion provides a new physical interpretation of how the nonlinear evolution of a single short-wave instability changes from weakly nonlinear saturation to evolution beyond the weakly nonlinear regime.

The above discussion suggests that larger linear growth rates of the elliptic and curvature instabilities lead to larger disturbance energies and stronger nonlinear effects, potentially resulting in significant deformation or breakdown of the helical vortex.
Such conditions can be achieved by larger Reynolds numbers and vortex core sizes for both instabilities, and by smaller reduced pitches for the elliptic instability in a typical parameter range.
These results provide useful insights for the control of helical tip vortices shed from wind turbines, for which the excitation of long-wave instabilities has already been considered as a means of promoting wake recovery \citep{brown_accelerated_2022, abraham_experimental_2023}.
The present findings may also be relevant to aircraft wake vortices, which are often idealized as a pair of columnar vortices and can likewise be destabilized by short-wave instabilities \citep{leweke_dynamics_2016}.
Since encounters with wake vortices can lead to loss of aircraft control, accelerating their decay through instabilities has also attracted attention \citep{edstrand_parallel_2018, zaccara_near_2024}.
How short-wave instabilities can be exploited for vortex control in these applications is an interesting direction for future work.

\begin{bmhead}[Acknowledgments.]
Numerical calculations were performed on AFI-NITY and AFI-NITY II at the Institute of Fluid Science, Tohoku University.
The corresponding author used ChatGPT (OpenAI) as an English-language-editing tool to improve grammar and sentence clarity.
All suggested revisions were carefully reviewed.
The scientific content and conclusions were not influenced by the use of this tool.
\end{bmhead}

\begin{bmhead}[Funding.]
This work was supported by JSPS KAKENHI 21H01242.
\end{bmhead}

\begin{bmhead}[Declaration of Interests.]
The authors report no conflict of interest.
\end{bmhead}

\begin{appen}
\normalsize

\section{Effect of numerical resolution}\label{appA}
The numerical scheme and spatial resolution employed here have been thoroughly investigated in \citet{hattori_numerical_2019, xu_numerical_2025}.
In particular, for a case with a higher Reynolds number and a smaller vortex core than those considered here, a minimum radial grid spacing of $\Delta r=5\times 10^{-3} R$ was shown to provide sufficient resolution.
To further assess the numerical resolution of the present simulations, we compare the present results with those obtained at a higher resolution, using $G_r\times G_\theta \times G_z =695\times 1280 \times 384$ and a minimum radial grid spacing $\Delta r=4.75\times 10^{-3} R$ and a time step$\Delta t=5\times 10^{-4} R^2/\Gamma$. 
For the elliptic-instability case $(\varepsilon,L/R)=(0.1,0.3)$, which has the smallest vortex core, the mode structures obtained from the linear stability analysis are shown in figures~\ref{fig:appenA}(a) and (b).
The radial structures obtained at the two resolutions are in close agreement, and the growth rate obtained at the higher resolution, $\sigma=6.60\times 10^{-2}$, is also in good agreement.
For the elliptic-instability case $(\varepsilon,L/R)=(0.2,0.3)$, in which the nonlinear evolution proceeds beyond the weakly nonlinear regime and  small-scale structures are generated, figure~\ref{fig:appenA}(c) compares the time evolution of the disturbance energies $E'_n$ obtained at the two resolutions.
The time evolutions of the unstable component and its higher harmonics are in good agreement between the two resolutions.
At the higher resolution, the maximum energy and the time at which it is attained are $E'_{1,\max}=2.057\times 10^{-3}$ and $t_{E'_{1,max}}=24.26$, respectively, compared with $E'_{1,\max}=2.061\times 10^{-3}$ and $t_{E'_{1,max}}=24.27$ at the present resolution.
The reasonable agreement is maintained even after nonlinear effects become appreciable, showing that the present resolution captures the nonlinear evolution for the present study.

\begin{figure}
  \centerline{\includegraphics[scale=.33]{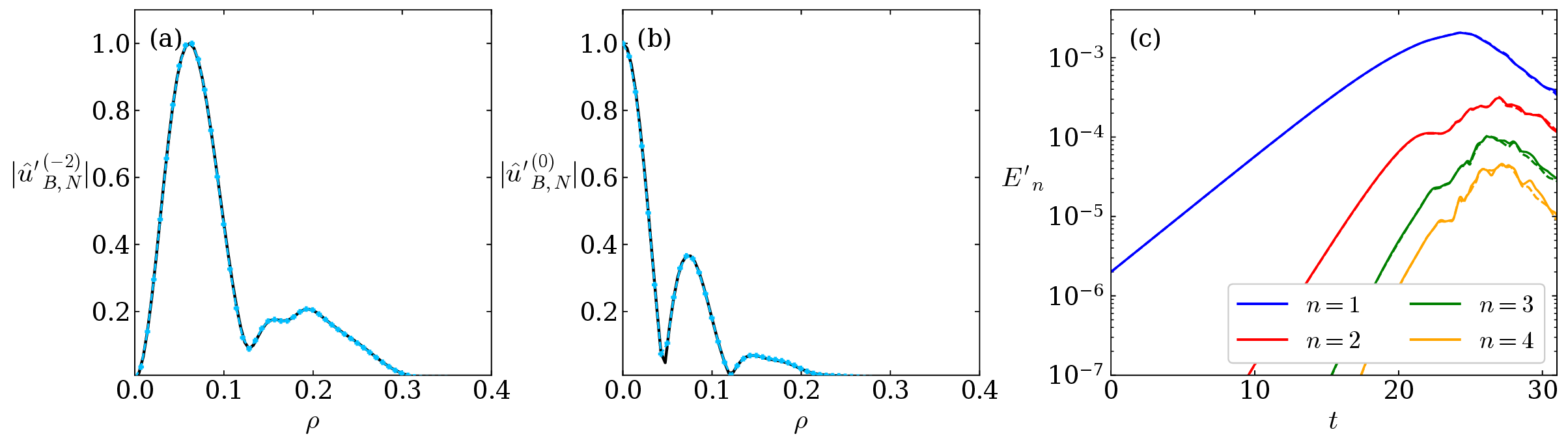}}
  \caption{Effect of numerical resolution. Normalized radial distributions of (a) $|\hat{u}'^{(-2)}_{B,N}|$ and (b) $|\hat{u}'^{(0)}_{B,N}|$ for the elliptic instability at $(\varepsilon,L/R)=(0.1,0.3)$, and (c) time evolution of the disturbance energies $E'_n$ for $(\varepsilon,L/R)=(0.2,0.3)$. Solid and dashed lines denote the higher and present resolutions, respectively; markers are added to the dashed lines in (a,b).}
\label{fig:appenA}
\end{figure}

\section{Effect of freezing the base flow in the nonlinear simulation}\label{appB}
For the weakly nonlinear regime, freezing the base flow provides a consistent framework for direct comparison with the weakly nonlinear analysis.
For evolution beyond the weakly nonlinear regime, \citet{schaeffer_nonlinear_2010} reported qualitatively similar dynamics when the viscous evolution of the base flow was allowed in a related configuration.
Following this, we examine the nonlinear evolution, with the base flow allowed to evolve under viscous effects.
We consider the elliptic-instability case $(\varepsilon,L/R)=(0.2,0.3)$, for which the nonlinear evolution proceeds beyond the weakly nonlinear regime.
In this calculation, the laboratory frame is employed, and a slightly larger initial disturbance amplitude is used than in the corresponding simulation.
The qualitative behavior remains similar to that obtained with the frozen base flow: the unstable component initially grows exponentially while generating higher harmonic components, and the helical vortex becomes substantially deformed near the time at which $E'_1$ reaches its maximum, as shown in figure~\ref{fig:appenB}.
The maximum energy of the unstable component is $E'_{1,\max}=1.998\times10^{-3}$ when the base flow is allowed to evolve viscously, compared with $E'_{1,\max}=2.061\times 10^{-3}$ in the calculation with the frozen case flow.
Thus, despite the differences in the numerical setting, the main features of the nonlinear evolution remain qualitatively similar.

\begin{figure}
  \centerline{\includegraphics[scale=.43]{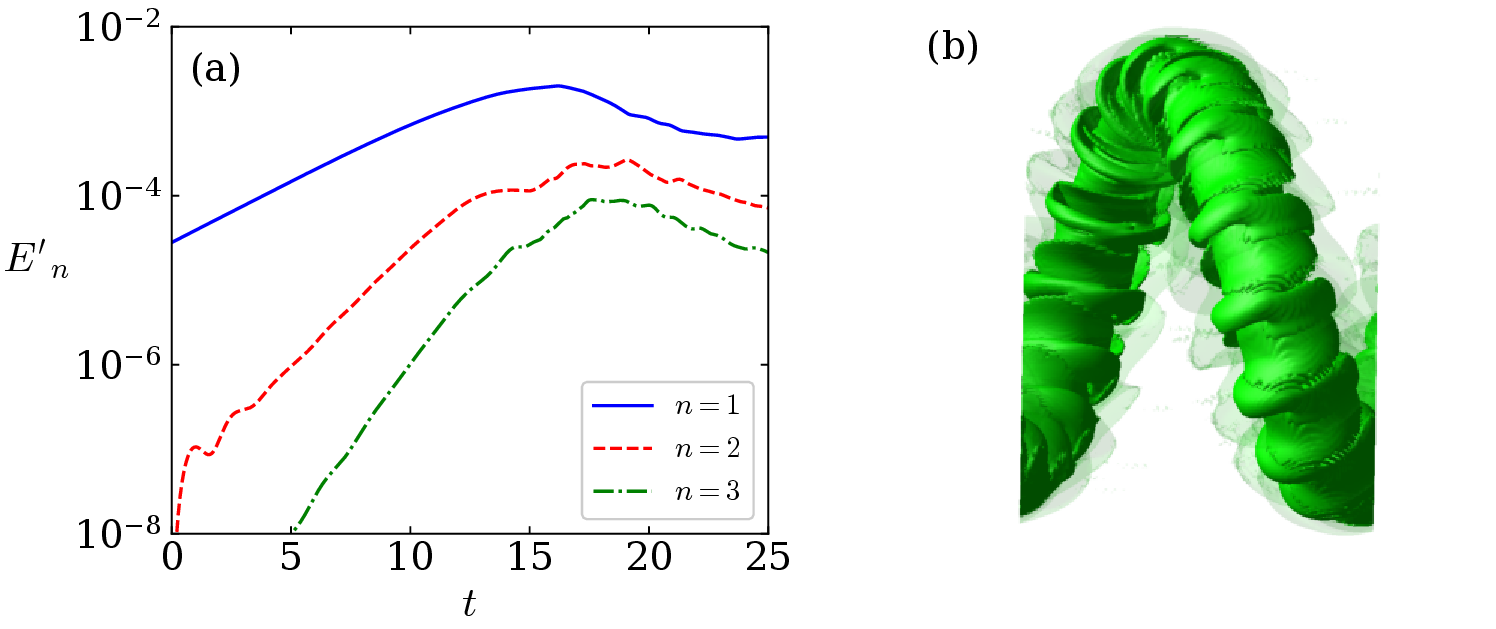}}
  \caption{(a)Time evolution of the disturbance energies $E'_n$ for the elliptic instability at $(\varepsilon,L/R)=(0.2,0.3)$, with the base flow allowed to evolve under viscous effects.
  (b)Isosurfaces of the vorticity magnitude $|\boldsymbol{\omega}|$ at $t=16.5$, when the energy of the unstable component is close to its maximum. The translucent and opaque isosurfaces correspond  to$|\boldsymbol{\omega}|/|\boldsymbol{\omega}_0|_{\max}=0.15$ and $|\boldsymbol{\omega}|/|\boldsymbol{\omega}_0|_{\max}=0.5$, respectively.}
\label{fig:appenB}
\end{figure}

\section{Weakly nonlinear analysis of short-wave instabilities}\label{appC}

\begin{table}
\centering

\begin{tabular}{cccccccc}
$c_{A0}$ &
$c_{B0}$ &
$c_{A1}$ &
$c_{A2}$ &
$c_{A3}$ &
$c_{B1}$ &
$c_{B2}$ &
$c_{B3}$ 
\\
$9.84\times 10^{-2}$ &
$-6.04\times 10^{-2}$ &
$0.187$ &
$1.26$ &
$0.989$ &
$10.72$ &
$-0.511$ &
$-1.33$ 
\\
\end{tabular}
\caption{Coefficients in the amplitude equations for the curvature instability $(-1,0;[2,4])$.}
\label{tab:appenC}

\end{table}

A common starting point for the weakly nonlinear analysis of the elliptic and curvature instabilities is an asymptotic expansion of $\boldsymbol{v}=[\boldsymbol{u},p]^{\mathrm{T}}$, where $p$ denotes the pressure, in the form

\begin{equation}\label{eq:expansion_v}
\begin{aligned}
\boldsymbol{v}
&= \sum_{i,j} \alpha^i \beta^j \boldsymbol{v}_{ij} \\
&= \boldsymbol{v}_{00}
+ \beta \boldsymbol{v}_{01}
+ \cdots
+ \alpha \boldsymbol{v}_{10}
+ \alpha^2 \boldsymbol{v}_{20}
+ \alpha^3 \boldsymbol{v}_{30}
+ \cdots
+ \alpha \beta \boldsymbol{v}_{11}
+ \alpha^2 \beta \boldsymbol{v}_{21}
+ \cdots .
\end{aligned}
\end{equation}
Here, $\alpha$ denotes a small parameter associated with the disturbance amplitude, while $\beta$ denotes the small parameter characterizing the departure of the base flow from a straight axisymmetric vortex.
The first and second subscripts of $\boldsymbol{v}_{ij}$ denote the orders in $\alpha$ and $\beta$, respectively.
For the elliptic instability, \citet{sipp_weakly_2000} considered a strained Lamb-Oseen vortex, with $\beta$ corresponding to the strain strength.
For the curvature instability, the same expansion was considered for a curved Batchelor vortex \citep{hattori_weakly_2026}, with $\beta=\varepsilon$.
The expansion of the base flow in $\varepsilon$, corresponding to the terms $\boldsymbol{v}_{0j}$, was previously obtained by \citet{blanco-rodriguez_internal_2015}.
In the curvature-instability formulation, a finite but small viscosity is retained to regularize the critical layer, while energy conservation is recovered in the inviscid limit.
In both cases, the leading-order disturbance $\boldsymbol{v}_{10}$ is expressed as a superposition of two resonant Kelvin waves.
For the curvature instability, for example, it can be written as

\begin{equation}
    \boldsymbol{v}_{10} = A\hat{\boldsymbol{v}}_A(\rho) e^{\mathrm{i}(-\psi+ks-\omega_A t)} +  B\hat{\boldsymbol{v}}_B(\rho) e^{\mathrm{i}(ks-\omega_B t)} + \mathrm{c.c.},
\end{equation}
where $k$ is the wavenumber along the vortex centerline, $\omega_A$ and $\omega_B$ are the frequencies of the corresponding Kelvin waves, and $s$ is the arc length along the vortex centerline: $z=sL/\sqrt{R^2+L^2}$ for a helical vortex along the center line helix.
Substitution of the asymptotic expansion into the incompressible Navier-Stokes equations yields equations for the higher-order components, with compatibility conditions arising at $O(\beta\alpha)$ and $O(\alpha^3)$ and a mean-flow correction appearing at $O(\beta\alpha^2)$.
These conditions introduce the slow time scale $\tau = \beta t=\alpha^2 t$ on which the amplitudes $A$, $B$ and $C$ evolve.
Neglecting linear damping at this stage, the resulting amplitude equations are

\begin{subequations}\label{eq:amplitude_equations_v2}
\begin{align}
\frac{\mathrm{d}A}{\mathrm{d}\tau}
&=
\mathrm{i} c_{A0} B
+
\left(
\mathrm{i}c_{A1}|A|^2
+
\mathrm{i}c_{A2}|B|^2
+
\mathrm{i}c_{A3}C
\right)A,
\label{eq:amp_A_v2}
\\
\frac{\mathrm{d}B}{\mathrm{d}\tau}
&=
\mathrm{i} c_{B0} A
+
\left(
\mathrm{i}c_{B1}|A|^2
+
\mathrm{i}c_{B2}|B|^2
+
\mathrm{i}c_{B3}C
\right)B,
\label{eq:amp_B_v2}
\\
\frac{\mathrm{d}C}{\mathrm{d}\tau}
&=
\mathrm{i}
\left(
A\overline{B}
-
\overline{A}B
\right).
\label{eq:amp_C_v2}
\end{align}
\end{subequations}
Including the leading-order disturbance amplitude $\alpha$ and damping terms into the amplitude equations, equations \eqref{eq:amplitude_equations} are obtained with $\kappa_A=\beta c_{A0}, \kappa_B=\beta c_{B0}, c_C=\beta, \mu_A=-\mathrm{Im}(\omega_A)-2\pi V_A/Re$, and $\mu_B=-2\pi V_B/Re$.
The values of $\mathrm{Im}(\omega_A),V_A$ and $V_B$ are taken from \citet{blanco-rodriguez_curvature_2017}.
Table \ref{tab:appenC} shows the coefficients used in \S~\ref{sec:discussion}, obtained in \cite{hattori_weakly_2026}.
The velocity component corresponding to $E'_{1,-1}$ is $A\hat{\boldsymbol{u}}_A$ at $O(\alpha)$, while that corresponding to $E'_{2,-2}$ is $A^2\hat{\boldsymbol{u}}_{AA}$ at $O(\alpha^2)$.
The latter arises from the second-order disturbance component $A^2 \hat{\boldsymbol{u}}_{AA} e^{\mathrm{i}(-2\psi+2ks-2\omega_A t)}$ contained in $\boldsymbol{v}_{20}$.

\end{appen}

\nocite{*}
\bibliographystyle{jfm_initials}
\bibliography{1JFM}

\end{document}